\documentclass[pra,twocolumn,amsmath,amssymb,superscriptaddress]{revtex4-1}
\usepackage{epsfig,amsmath}
\usepackage{subfigure}
\usepackage{graphicx}
\usepackage{dcolumn}
\usepackage{stmaryrd}
\usepackage{mathrsfs}
\usepackage{pifont}
\usepackage{amsthm}
\usepackage{amssymb}
\usepackage{bm}
\usepackage{latexsym}
\usepackage[colorlinks=true,linkcolor=blue,citecolor=blue]{hyperref}
\usepackage{color}
\usepackage{epstopdf}

\usepackage{booktabs}
\usepackage{threeparttable}
\usepackage{multirow}
\usepackage{array}
\usepackage{mathdots}

\begin{document}

\title{Criterion for qubit-assisted quantum metrology exhibiting Heisenberg scaling}
\author{Peng Chen}
\affiliation{School of Physics, Zhejiang University, Hangzhou 310027, Zhejiang, China}

\author{Jun Jing}
\email{Contact author: jingjun@zju.edu.cn}
\affiliation{School of Physics, Zhejiang University, Hangzhou 310027, Zhejiang, China}

\date{\today}

\begin{abstract}
We study in this work the metrology precision of a probe system coupled to an ancillary qubit. Restricting the probe-qubit coupling along no more than two orthogonal directions is found to be a sufficient criterion for the effective dynamical generator of the composite system to exhibit Heisenberg scaling in precision. Under this coupling criterion, the quantum Fisher information (QFI) about the to-be-estimated parameter becomes the mean square of the effective generator with respect to the initial state of the composite system. Our criterion is justified in two distinct systems. For a bosonic probe, QFI about the displacement estimation is found to be proportional to the average excitation number of probe. It is seemingly counterintuitive that quantum metrology sensitivity can be enhanced by increasing the probe system's temperature. For a spin-ensemble probe, QFI about both the rotation phase and magnetic field can exhibit a quadratic dependence on the probe-spin number, even when the probe is initially in a mixed state. Our protocol actually manufactures the metrologically relevant states to manifest a Heisenberg scaling behavior through qubit-probe interaction. At the input, it relieves the probe of the optimization to the so-called resource states, e.g., the squeezed state or the Greenberger-Horne-Zeilinger (GHZ) state, in sharp contrast to the prior preparation-then-encoding paradigm.
\end{abstract}

\maketitle

\section{Introduction}

Quantum metrology represents one of the most promising applications of quantum technologies~\cite{giovannetti2006quantum,budker2007optical,paris2009quantum,degen2017quantum}. It shows a remarkable potential in a wide range of applications, including atomic clock~\cite{Katori2011optical,Ludlow2015optical}, gravitational wave detection~\cite{Caves1981quantum}, biological sensing~\cite{Taylor2016quantum,Mauranyapin2017evanescent}, and magnetometry~\cite{Jones2009magnetic}. In all of these scenarios, the parameter to be estimated is encoded into the probe state through a parameter generator. Using classical states, the generator structure defines the standard quantum limit (SQL)~\cite{giovannetti2004quantum,toth2014quantum,nawrocki2015introduction} about the achievable precision. When certain metrologically relevant resources are introduced to the probe system, the ultimate precision can surpass the SQL and approach the Heisenberg limit (HL)~\cite{giovannetti2011advances,pezze2018quantum,polino2020photonic,ye2024essay,demille2024quantum}.

Parametric estimation in metrology protocols is conventionally quantified by the scaling of standard deviation with the number of probe subunits. For displacement estimation in photonic systems~\cite{genoni2013optimal,duivenvoorden2017single,bond2025optimal}, the SQL is defined in reference to coherent states, yielding a state-independent standard deviation of $1/2$. By instead using squeezed states, the precision can approach the HL $1/(2\sqrt{4\bar{n}+1})$ with $\bar{n}$ the average photon number of the probe mode~\cite{degen2017quantum,wolf2019motional,valahu2025quantum}. A trapped-ion platform employing a single-photon Fock state $|n=1\rangle$~\cite{wolf2019motional} demonstrated a precision exceeding the SQL by achieving a measurement uncertainty of $0.43<1/2$. For phase estimation in photonic systems~\cite{higgins2007entanglement,resch2007time,xu2013phase}, the SQL behaves as $1/\sqrt{\bar{n}}$, whereas the HL scales as $1/\bar{n}$. A measurement precision approaching the HL has been confirmed by using a six-photon NOON state with a fidelity of $90\pm2\%$~\cite{resch2007time}. For rotation-phase estimation in atomic systems~\cite{gross2010nonlinear,d2024atom,shaw2024multi}, the SQL is defined by spin coherent states as $1/\sqrt{N}$ and HL scales as $1/N$ with respect to the atomic number $N$. A spin-squeezed state of $120$ atoms has achieved an estimation uncertainty of $1/27.39$~\cite{gross2010nonlinear}, outperforming the corresponding SQL magnitude $1/\sqrt{120}$, and yet remaining far from the HL due to the detection noise~\cite{esteve2006observations,esteve2008squeezing}.

Recently, coupling the probe system to an ancillary system has been regarded as an unconventional resource to relax or even replace initial requirements in various metrologies~\cite{boixo2007generalized,ritboon2022sequential,
yang2022variational,xia2023nanoradian,fan2024achieving,chen2024qubit,chen2025achieving}, such as entanglement or squeezing. The motional degree of freedom of a trapped ion (a bosonic-mode probe) is coupled to its internal electronic levels (an ancillary qubit) through a Jaynes-Cummings (JC) interaction~\cite{ritboon2022sequential}. The information about the displacement encoded in the probe system can be indirectly extracted by measurements on the ancillary qubit and the precision allows one to approach the HL. For a probe-frequency estimation assisted by a coupled qubit~\cite{fan2024achieving}, the parameter information can be extracted via measurements on the ancillary qubit after tracing out the probe state. By properly tailoring the interaction Hamiltonian, the time points for measurements, and the coupling strength, the estimation precision can periodically attain the Heisenberg scaling in terms of probe size and time. In a protocol for estimating the rotation phase~\cite{chen2024qubit} in a spin ensemble, a time-reversal strategy can be realized by a coupled ancillary qubit with a ZZ interaction, enabling unitary transformations along the forward- and reverse-time directions. Then quantum Fisher information (QFI) can approach the HL regardless of the purity of the probe state, which takes advantage over the prior protocols. Nevertheless, it is still unknown how to determine the scope of probe-ancilla coupling, in which metrology can realize the Heisenberg scaling.

Our work aims to fill this gap by establishing a sufficient coupling criterion to identify in the qubit-assisted paradigm whether or not that QFI associated with the encoded parameter can {\em always} attain its upperbound in metrologies, i.e., the mean square of the effective dynamical generator with respect to {\em an arbitrary initial state}. It does not mean that we would propose a resource-free metrology. Instead, we are working in a paradigm that a metrologically relevant state can be generated by the system evolution of a constrained duration. The paradigm has been well steered experimentally and theoretically~\cite{mccormick2019quantum,deng2024quantum,pan2025realization}. For a bosonic probe, it is found that QFI about the displacement estimation is proportional to the average photon number $\bar{n}$. This result indicates that the estimation precision can be enhanced by increasing the temperature of the bosonic mode. For a spin-ensemble probe prepared as a mixed state, e.g., a thermal state, it can also attain an asymptotic Heisenberg-scaling behavior with respect to the probe spin number $N$ in rotation-phase estimation. In addition, our criterion applies to weak-field magnetometry, where the parametric encoding is embedded in the joint evolution of the composite system rather than where the evolution and encoding are separable.

The rest of this work is structured as follows. In Sec.~\ref{section criterion}, we derive a general coupling criterion for qubit-assisted metrology. Following the criterion, we investigate the optimized protocols in qubit-bosonic and qubit-spin-ensemble systems in Secs.~\ref{subsection probe bosonic system} and \ref{subsection probe spin system}, respectively, where the encoding and evolution are separable. For each system, we analyze the effects from the parametric deviation and external decoherence on QFI. In Sec.~\ref{section entangle encoding}, we perform our metrology on the weak-field magnetometry in both idea and nonideal situations, where the parametric encoding is embedded in time evolution. The entire work is summarized in Sec.~\ref{section conclusion}. Appendix~\ref{appendix evolution operator} presents the detailed derivations about the joint time-evolution operators for both qubit-bosonic and qubit-spin-ensemble systems. Explicit solutions of our criterion for these systems are given in Appendix~\ref{appendix solution for criteria}. Appendix~\ref{appendix classical Fisher information} provides a practical measurement strategy in which the classical Fisher information (CFI) saturates the reported QFI.

\section{Criterion for quantum metrology assisted by coupled qubit}\label{section criterion}

We start by briefly recalling the quantum Fisher information for an ancilla-free probe. When the probe $\rho_P$ experiences a unitary parametrization process $U_{\theta}$ with a to-be-estimated parameter $\theta$, the QFI is defined as $F_Q\equiv{\rm Tr}(\rho_{\theta}L^2)={\rm Tr}(U_{\theta}\rho_PU_{\theta}^{\dagger}L^2)$, where $L$ is the symmetric logarithmic derivative operator~\cite{helstrom1969quantum,holevo2011probabilistic} defined by   $\partial_{\theta}\rho_{\theta}=(L\rho_{\theta}+\rho_{\theta}L)/2$. $U_{\theta}$ can be either implemented in a time evolution or in an independent quench. Assuming the spectral decomposition of the probe density matrix is $\rho_P=\sum_{k=1}^dp_k|\psi_k\rangle\langle\psi_k|$ with $p_k\geq0$, $\langle\psi_j|\psi_k\rangle=\delta_{kj}$, and $\sum_{k=1}^dp_k=1$, the QFI can be expressed as~\cite{braunstein1994statistical,braunstein1996generalized,zhang2013quantum,liu2014quantum}
\begin{equation}\label{QFI for probe state}
F_Q=\sum_{k=1}^d4p_k\langle\mathcal{H}^2\rangle_k
-\sum_{k,j=1}^d\frac{8p_kp_j}{p_k+p_j}|\langle\mathcal{H}\rangle_{kj}|^2.
\end{equation}
Here $\mathcal{H}\equiv iU_\theta^\dagger(\partial_\theta U_\theta)$ is the effective dynamical generator and $\langle\cdot\rangle_{kj}\equiv\langle\psi_k|\cdot|\psi_j\rangle$ that is simplified to $\langle\cdot\rangle_k$ when $k=j$. One can immediately find that the QFI is a difference between two positive terms. The first term is crucial to achieve the Heisenberg scaling. A practical idea to enhance the QFI is thus to minimize the second term, or more generally and precisely, to minimize the magnitude of $|\langle\mathcal{H}\rangle_{kj}|$. Conventionally, by introducing nonclassical resources~\cite{giovannetti2011advances,pezze2018quantum}, such as entanglement or squeezing, to the probe system, the second term in Eq.~(\ref{QFI for probe state}) can be reduced, thereby significantly enhancing the QFI. We show that is not necessary at the input.

Consider a probe system coupled to an ancillary spin-$1/2$ and assume that the two components are initially separable, i.e., the input state of the composite system is $\rho_P\otimes\rho_A$ with arbitrary $\rho_P$ and $\rho_A=|\varphi\rangle\langle\varphi|$. Then Eq.~(\ref{QFI for probe state}) becomes
\begin{equation}\label{QFI for probe state and ancillary qubit}
F_Q=\sum_{k=1}^d4p_k\langle\varphi|\langle\mathcal{H}^2\rangle_k|\varphi\rangle
-\sum_{k,j=1}^d\frac{8p_kp_j}{p_k+p_j}
|\langle\varphi|\langle\mathcal{H}\rangle_{kj}|\varphi\rangle|^2,
\end{equation}
where the effective dynamical generator $\mathcal{H}$ for the composite system can be expanded by the identity and Pauli operators of the ancillary qubit, i.e.,
\begin{equation}\label{effective phase generator}
\mathcal{H}=iU_\theta^\dagger(\partial_\theta U_\theta)=M+\vec{\boldsymbol r}\cdot\vec{\boldsymbol\sigma}=M+X\sigma_x+Y\sigma_y+Z\sigma_z.
\end{equation}
Here $\vec{\boldsymbol\sigma}=(\sigma_x,\sigma_y,\sigma_z)$ is the operator vector with Pauli matrices and $\vec{\boldsymbol r}\equiv(X,Y,Z)$. $M={\rm Tr}_A(\mathcal{H})/2$, $X={\rm Tr}_A(\mathcal{H}\sigma_x)/2$, $Y={\rm Tr}_A(\mathcal{H}\sigma_y)/2$, and $Z={\rm Tr}_A(\mathcal{H}\sigma_z)/2$ are the probe operators with ${\rm Tr}_A$ the partial trace over the ancillary qubit. Inserting Eq.~(\ref{effective phase generator}) to Eq.~(\ref{QFI for probe state and ancillary qubit}), we have
\begin{equation}\label{QFI for probe state and ancillary qubit in terms of pauli matrix}
\begin{aligned}
F_Q=&\sum_{k=1}^d4p_k\left(\langle M^2+Z^2+X^2+Y^2\rangle_k+\langle\vec{\boldsymbol R}\rangle_k\cdot\langle\vec{\boldsymbol\sigma}\rangle_\varphi\right)\\
-&\sum_{k,j=1}^d\frac{8p_kp_j}{p_k+p_j}|\langle M\rangle_{kj}+\langle\vec{\boldsymbol r}\rangle_{kj}\cdot\langle\vec{\boldsymbol\sigma}\rangle_\varphi|^2,
\end{aligned}
\end{equation}
where $\langle\cdot\rangle_\varphi\equiv\langle\varphi|\cdot|\varphi\rangle$ and $\vec{\boldsymbol R}\equiv(\{M,X\}+i[Y,Z], \{M,Y\}+i[Z,X], \{M,Z\}+i[X,Y])$. In the absence of qubit-probe coupling, Eq.~(\ref{QFI for probe state and ancillary qubit}) reduces to Eq.~(\ref{QFI for probe state}). Otherwise we have one more degree of freedom to enhance the QFI by optimizing the qubit state $|\varphi\rangle$. From this perspective, the upperbound of the QFI in Eq.~(\ref{QFI for probe state and ancillary qubit in terms of pauli matrix}) is attained under the following two conditions:
\begin{subequations}
\begin{align}\label{condition for the first term of QFI}
&\langle\vec{\boldsymbol R}\rangle_k\cdot\langle\vec{\boldsymbol\sigma}\rangle_\varphi=|\langle\vec{\boldsymbol R}\rangle_k|,\\ \label{condition for the second term of QFI}
&\langle M\rangle_{kj}+\langle\vec{\boldsymbol r}\rangle_{kj}\cdot\langle\vec{\boldsymbol\sigma}\rangle_\varphi=0.
\end{align}
\end{subequations}
The first condition is based on the Cauchy-Schwarz inequality and the second one is to cancel the negative contribution from the second term in Eq.~(\ref{QFI for probe state and ancillary qubit in terms of pauli matrix}). A sufficient criterion for both Eqs.~(\ref{condition for the first term of QFI}) and (\ref{condition for the second term of QFI}) reads
\begin{subequations}
\begin{align}\label{criterion about M}
M=0,\\  \label{criterion about XYZ}
X=0 \quad  {\rm or} \quad Y=0 \quad  {\rm or} \quad Z=0.
\end{align}
\end{subequations}
If $M=0$ and $X=0$, then $\vec{\boldsymbol R}=(i[Y, Z], 0, 0)$. Consequently, Eqs.~(\ref{condition for the first term of QFI}) and (\ref{condition for the second term of QFI}) are reduced to
\begin{subequations}
\begin{align}\label{condition for the first term of QFI example}
& i\langle[Y,Z]\rangle_k\langle\sigma_x\rangle_\varphi=|\langle i[Y,Z]\rangle_k|,\\
\label{condition for the second term of QFI example}
& \langle Z\rangle_{kj}\langle\sigma_z\rangle_\varphi+\langle Y\rangle_{kj}\langle\sigma_y\rangle_\varphi=0,
\end{align}
\end{subequations}
respectively. It is straightforward to check that for an arbitrary probe state, both Eqs.~(\ref{condition for the first term of QFI example}) and (\ref{condition for the second term of QFI example}) are valid as long as the qubit is prepared as a balanced superposition $|\varphi\rangle=|\pm\rangle\equiv(|e\rangle\pm|g\rangle)/\sqrt{2}$ with $|e\rangle$ and $|g\rangle$ the excited and ground states, respectively. Consequently, the QFI in Eq.~(\ref{QFI for probe state and ancillary qubit in terms of pauli matrix}) becomes a mean square of the effective generator with respect to the initial state of the composite system,
\begin{equation}\label{QFI with optimized qubit}
\begin{aligned}
F_Q&=4{\rm Tr}\left(\mathcal H^2\rho_P\otimes\rho_A\right)\\
&=\sum_{k=1}^d4p_k\left(\langle Z^2+Y^2\rangle_k+|\langle i[Y,Z]\rangle_k|\right)\\
&=4{\rm Tr}[(Z-iY)(Z+iY)\rho_P].
\end{aligned}
\end{equation}
Note that the criterion in Eqs.~(\ref{criterion about M}) and (\ref{criterion about XYZ}) is sufficient for the arbitrary initial state of the probe yet not necessary to hold Eqs.~(\ref{condition for the first term of QFI}) and (\ref{condition for the second term of QFI}). For example, for estimating the rotation phase in a spin ensemble, QFI can attain the HL through being the mean square of the effective generator if one employs a GHZ-like state~\cite{chen2024qubit}. In that case, $M$ is found to be nonvanishing.

\section{Qubit-assisted metrology under separable encoding and evolution}\label{section separable encoding}

Qubit-assisted metrologies can be generally divided into two scenarios, due to whether or not the joint free-evolution of the composite system and parametric encoding are simultaneously implemented. This section devotes to the scenario that encoding and evolution are separable, which means the entire evolution operator is described by
\begin{equation}\label{parametrization process in atomic system}
U_{\theta}=R(\theta)U(t)=e^{-i\theta G}e^{-iHt},
\end{equation}
where $\theta$ is the to-be-estimated parameter imprinted via the parameter generator $G$, $H$ is the full Hamiltonian, and $t$ is the duration of a joint evolution.

\subsection{Qubit-bosonic system}\label{subsection probe bosonic system}

When the probe system is a single cavity mode, the full Hamiltonian is generally given by $(\hbar\equiv1)$
\begin{equation}\label{qubit boson hamiltionian}
H=\omega_Pa^\dagger a+\omega _A\sigma_z+g_za^\dagger a\sigma_z+g\left(a^\dagger\sigma_-+a\sigma_+\right),
\end{equation}
where $a^\dagger(a)$ indicates the bosonic creation (annihilation) operator and $\sigma_\pm=(\sigma_x\pm i\sigma_y)/2$ are the raising and lowering operators of the ancillary qubit. $\omega_P$ and $\omega_A$ denote the frequencies of the bosonic mode and the ancillary spin, respectively. $g$ and $g_z$ characterize their exchange and dispersive or longitudinal coupling strengths, respectively. When $g_z=0$, the composite system reduces to a JC model. The time-evolution operator $U(t)$ driven by a general $H$ is given by Eq.~(\ref{expression full evolution operator}).

In the bosonic probe system, the encoding operator $R(\theta)$ performed on the probe is typically a displacement operator in phase space along a certain direction~\cite{genoni2013optimal,duivenvoorden2017single,bond2025optimal}. Then the phase generator is reasonable to be
\begin{equation}\label{boson generator}
G=\frac{e^{-i\xi}a+e^{i\xi}a^\dagger}{\sqrt2},
\end{equation}
with $\xi$ a real number. Using Eqs.~(\ref{parametrization process in atomic system}) and (\ref{boson generator}), the four components of the effective dynamical generator read
\begin{subequations}
\begin{align}\label{boson matrix elements M}
M&=\frac{e^{-i\xi}{\rm Tr}_A\left(e^{iHt}ae^{-iHt}\right)+{\rm H.c.}}{2\sqrt2},\\\label{boson matrix elements XYZ}
X,Y,Z&=\frac{e^{-i\xi}{\rm Tr}_A\left(e^{iHt}ae^{-iHt}\sigma_{x,y,z}\right)+{\rm H.c.}}{2\sqrt2}.
\end{align}
\end{subequations}
Substituting the evolution operator~(\ref{expression full evolution operator}) to Eq.~(\ref{boson matrix elements M}), one can find that $M=0$, i.e., the criterion in Eq.~(\ref{criterion about M}) can be deduced by
\begin{equation}\label{expression criterion M=0}
e^{i\Omega(n+1)t}+e^{i\Omega(n-1)t}=0,
\end{equation}
with
\begin{equation}\label{boson frequency_definition}
\Omega(n)\equiv\sqrt{g_z^2n^2+(g^2+g_z^2+2g_z\Delta)n+g^2+\left(\frac{g_z}{2}+\Delta\right)^2},
\end{equation}
where $n$ is a nonnegative integer and $\Delta=\omega_A-\omega_P/2$ is the detuning between the qubit and bosonic mode. Using the integer-valued polynomials~\cite{cahen1997integer}, Eq.~(\ref{expression criterion M=0}) is equivalent to
\begin{equation}\label{solution for criteria boson M}
\sum_{k=0}^{k_{\rm max}}\sum_{\substack{l=0\wedge\\k-l={\rm odd}}}^{k-1}C_k^ld_kn^l=\frac{\pi}{t}\left(\sum_{k=0}^{k_{\rm max}-1}\sum_{l=0}^ka_k\frac{s(k,l)}{k!}n^l+\frac{1}{2}\right),
\end{equation}
where $C_k^l=k!/[l!(k-l)!]$ denotes the binomial coefficient, $a_k$ is an arbitrary integer, $s(k,l)$ indicates the Stirling numbers of the first kind, and $d_k$ is the $k$th coefficient in the Taylor expansion of $\Omega(n)$ with respect to $n$, i.e.,
\begin{equation}\label{boson frequency}
\Omega(n)=\sum_{k=0}^{k_m}d_kn^k,
\end{equation}
up to the order $k_m$.

In addition, it is found that the second part of criterion $X=0$ or $Y=0$ in Eq.~(\ref{criterion about XYZ}) can be deduced by
\begin{equation}\label{expression criterion XY=0}
{\rm Im}\left(e^{i\Omega(n)t}e^{-i\Omega(n+1)t}\right)=0,
\end{equation}
which is equivalent to
\begin{equation}\label{solution for criteria boson XY}
\sum_{k=0}^{k_{\rm max}}\sum_{\substack{l=0\wedge\\k-l={\rm even}}}^{k-2}C_k^ld_kn^l=\frac{\pi}{t}\left(\sum_{k=0}^{k_{\rm max}-2}\sum_{l=0}^kb_k\frac{s(k,l)}{k!}n^l+\frac{1}{2}\right),
\end{equation}
with $b_k$ an arbitrary integer. The derivation details about Eqs.~(\ref{solution for criteria boson M}) and (\ref{solution for criteria boson XY}) can be found in Appendix~\ref{appendix solution for criteria}.

Comparing Eq.~(\ref{solution for criteria boson M}) to Eq.~(\ref{boson frequency}), one can find that $k_m=k_{\rm max}$. A larger $k_{\rm max}$ yields a more accurate description about Eq.~(\ref{boson frequency}). For example, we set $k_{\rm max}=k_m=1$ so that Eq.~(\ref{boson frequency}) is simplified to
\begin{equation}\label{boson coefficients}
\Omega(n)=\frac{g^2+g_z(g_z+2\Delta)}{\sqrt{4g^2+(g_z+2\Delta)^2}}n+\sqrt{g^2+\left(\frac{g_z}{2}+\Delta\right)^2}.
\end{equation}
In this case, Eq.~(\ref{solution for criteria boson M}) reduces to
\begin{equation}\label{solution for criteria boson simple example}
\frac{t}{\pi}\frac{g^2+g_z(g_z+2\Delta)}{\sqrt{4g^2+(g_z+2\Delta)^2}}=a_0+\frac{1}{2},
\end{equation}
and Eq.~(\ref{solution for criteria boson XY}) is simultaneously valid. Equation~(\ref{solution for criteria boson simple example}) constitutes a constraint for $t$, $\Delta$, $g$, and $g_z$.

Upon the conditions in Eqs.~(\ref{expression criterion M=0}) and (\ref{expression criterion XY=0}), the effective dynamical generator in Eq.~(\ref{effective phase generator}) is found to be
\begin{equation}\label{boson dynamical generator}
\mathcal{H}=Z\sigma_z=\frac{i}{\sqrt2}\left[e^{i(\xi+\omega_Pt)}a^\dagger-e^{-i(\xi+\omega_Pt)}a\right]\sigma_z.
\end{equation}
When the initial ancillary qubit is prepared in $|+\rangle$, the QFI in Eq.~(\ref{QFI for probe state and ancillary qubit}) becomes
\begin{equation}\label{boson qfi}
F_Q=4{\rm Tr}\left(Z^2\rho_P\right).
\end{equation}
The Heisenberg scaling with respect to the amplitude of the probe mode, $F_Q=8\alpha^2+2$, is exactly attained when the probe is prepared in coherent states $|\pm\alpha e^{i(\xi+\omega_Pt-\pi/2)}\rangle$ or an arbitrary mixed or superposition state over them. Here $|\pm\alpha e^{i(\xi+\omega_Pt-\pi/2)}\rangle$ are eigenstates of the optimized operator $Z$ with $\alpha$ a positive number. Note that once the conditions in Eqs.~(\ref{expression criterion M=0}) and (\ref{expression criterion XY=0}) are satisfied, exemplified by Eq.~(\ref{solution for criteria boson simple example}), the quantum Fisher information in Eq.~(\ref{boson qfi}) is irrelevant to the specific magnitudes of $g$, $g_z$, $\Delta$, and $t$.

In a practical scenario, however, it is generally difficult to control these systematic parameters with perfect accuracy. According to Eq.~(\ref{solution for criteria boson simple example}), one can find that the effect of imprecise control over parameters on the effective dynamical generator and hence the QFI, can be attributed to the integer coefficient $a_0$. If $a_0\rightarrow a'_0=a_0+\delta a_0$, where $|\delta a_0|\ll1$ indicates the deviation caused by any parametric error, the condition is modified to be
\begin{equation}
\frac{t}{\pi}\frac{g^2+g_z(g_z+2\Delta)}{\sqrt{4g^2+(g_z+2\Delta)^2}}=a_0+\delta a_0+\frac{1}{2},
\end{equation}
so that the Rabi frequency $\Omega(n)$ in Eq.~(19) varies as
\begin{equation}\label{Rabi frequency bosonic}
\Omega(n)=\left(a_0+\frac{1}{2}+\delta a_0\right)\frac{n\pi}{t}+\sqrt{g^2+\left(\frac{g_z}{2}+\Delta\right)^2}.
\end{equation}
Substituting Eqs.~(\ref{Rabi frequency bosonic}) and (\ref{expression full evolution operator}) into Eqs.~(\ref{boson matrix elements M}) and (\ref{boson matrix elements XYZ}), one can find that the four components of the effective dynamical generator read $X\approx Y\approx0$ and
\begin{subequations}
\begin{align} \label{M bosonic}
M&\approx(-1)^{a_0+1}\frac{e^{-i(\xi+\omega_Pt)}a
+e^{i(\xi+\omega_Pt)}a^\dagger}{\sqrt{2}}\pi\delta a_0, \\ \label{Z bosonic}
Z&\approx(-1)^{a_0+1}\frac{e^{-i(\xi+\omega_Pt)}a
-e^{i(\xi+\omega_Pt)}a^\dagger}{i\sqrt{2}}\left[1-\frac{(\pi\delta a_0)^2}{2}\right],
\end{align}
\end{subequations}
up to the leading order of $\delta a_0$. Upon the initial state of the composite system $|\alpha e^{i(\xi+\omega_Pt-\pi/2)}\rangle\otimes|+\rangle$ , the QFI in Eq.~(\ref{QFI for probe state and ancillary qubit in terms of pauli matrix}) reduces to
\begin{equation}
F_Q\approx8\alpha^2\left[1-(\pi\delta a_0)^2\right]+2,
\end{equation}
up to the second order of $\delta a_0$. This result indicates that our protocol as well as the optimized condition in Eq.~(\ref{solution for criteria boson simple example}) is insensitive to small parametric deviations.

\begin{figure}[htbp]
\begin{centering}
\includegraphics[width=0.8\linewidth]{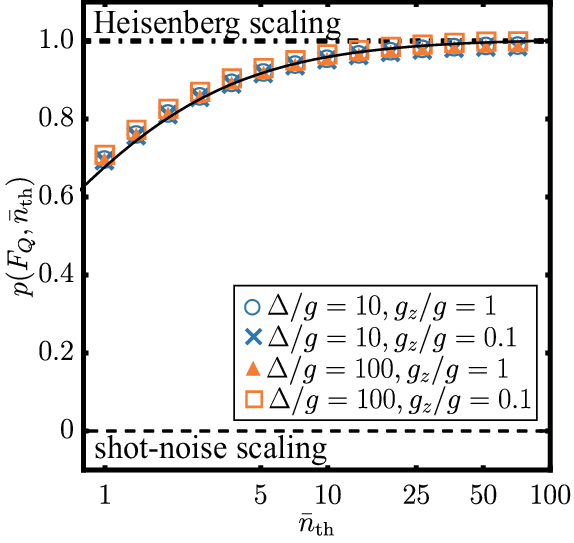}
\caption{Scaling exponent $p(F_Q, \bar{n}_{\rm th})$ of the QFI about displacement estimation as a function of the average photon number $\bar{n}_{\rm th}$ for a mixed state $\rho_b^{\rm th}$. The solid line is the analytical result in Eq.~(\ref{boson qfi thermal state}). The black dashed line and the black dot-dashed line indicate the Heisenberg and shot-noise scalings, respectively. Here $\rho_A=|+\rangle\langle+|$.}\label{QFI boson thermal state}
\end{centering}
\end{figure}

The Heisenberg scaling can be asymptotically attained even if the probe loses its coherence to become a mixed state, e.g., the thermal state $\rho_P=\rho_b^{\rm th}=e^{-\beta a^\dagger a}/Z_\beta$, where $Z_\beta={\rm Tr}[\exp(-\beta a^\dagger a)]$ is the partition function and $\beta$ is the inverse temperature. By Eq.~(\ref{boson qfi}), we have
\begin{equation}\label{boson qfi thermal state}
\begin{aligned}
F_Q&=2\frac{\sum_{n=0}^\infty(2n+1)\exp(-n\beta)}{\sum_{n=0}^\infty\exp(-n\beta)}\\
&=2\frac{e^{\beta}+1}{e^{\beta}-1}=4\bar{n}_{\rm th}+2,
\end{aligned}
\end{equation}
where $\bar{n}_{\rm th}=1/(e^\beta-1)$ is the average photon number of the initial state. Clearly, $F_Q\rightarrow\infty$ when $\bar{n}_{\rm th}\rightarrow\infty$. The seemingly counterintuitive behavior in Eq.~(\ref{boson qfi thermal state}) can be confirmed by the numerical simulation over the scaling exponent $p(F_Q, \bar{n}_{\rm th})\equiv d\ln F_Q/d\ln \bar{n}_{\rm th}$ in Fig.~\ref{QFI boson thermal state}, under various $\Delta$ and $g_z$ in Eq.~(\ref{solution for criteria boson simple example}) that yield a relevant joint evolution time $t$. QFI is verified to be insensitive to specific $g$, $g_z$, $\Delta$, and $t$ given by Eq.~(\ref{solution for criteria boson simple example}). And our results indeed exhibit an approximate Heisenberg scaling behavior for a sufficiently large $\bar{n}_{\rm th}$. Formally, it is a half of that obtained by the squeezed state~\cite{degen2017quantum}. When $\bar{n}_{\rm th}\geq 25$, it is hard to distinguish the numerical simulation from the Heisenberg scaling, since $p(F_Q,n_{\rm th})\approx0.98$.

Next, we consider a nonunitary evolution of the composite system, e.g., in the presence of a dissipative environment. The evolution of the full density matrix can be described by the Lindblad master equation
\begin{equation}\label{masterEquation of bosonic system}
\dot\rho(t)=-i[H, \rho(t)]+\kappa\mathcal L[a]\rho(t)+\gamma_P\mathcal L[a^\dagger a]\rho(t)+\gamma_A\mathcal L[\sigma_z]\rho(t),
\end{equation}
where $\kappa$ and $\gamma_P$ denote the decay and dephasing rates of the probe system, respectively, and $\gamma_A$ denotes the qubit dephasing rate. The Lindblad superoperator $\mathcal L$ is defined as $\mathcal L[O]\rho\equiv O\rho O^\dagger-\frac{1}{2}(O^\dagger O\rho+\rho O^\dagger O)$, where $O=a$, $a^\dagger a$, or $\sigma_z$.

\begin{figure}[htbp]
\begin{centering}
\includegraphics[width=0.8\linewidth]{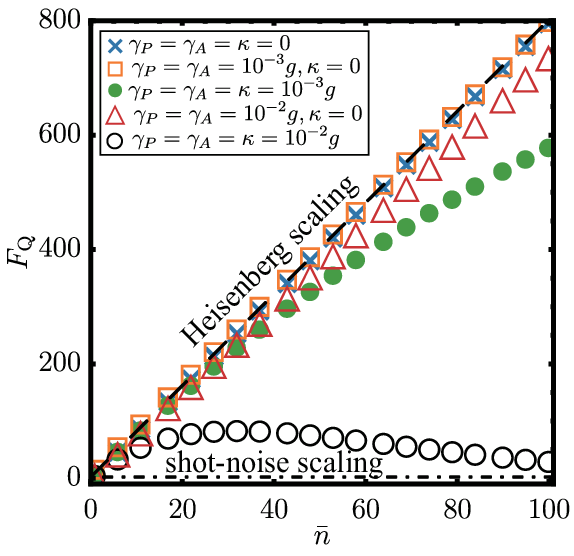}
\caption{QFI as a function of the average excitation number $\bar{n}$ for the probe in an optimized coherent state under various decoherence rates. The black dashed line and the black dot-dashed line indicate the Heisenberg scaling and the shot-noise scaling, respectively. Here $\rho_A=|+\rangle\langle+|$, $\Delta=100g$, $g_z=g$, and $t\approx\pi/(2g)$.}\label{QFI_dissipative_boson}
\end{centering}
\end{figure}

In Fig.~\ref{QFI_dissipative_boson}, we present QFI under various decay and dephasing rates, where the probe and qubit are initialized as the optimized coherent state $|\alpha e^{i(\xi+\omega_Pt-\pi/2)}\rangle$ and pure state $|+\rangle$. Without loss of generality, we choose the parameters $\Delta=100g$, $g_z=g$, and $a_0=0$, which yield $t\approx\pi/(2g)$ by Eq.~(\ref{solution for criteria boson simple example}). In the absence of decoherence, i.e., $\gamma_P=\gamma_A=\kappa=0$, it is found that the QFI can exactly follow the Heisenberg scaling $F_Q=8\bar n+2$, which is in agreement with Eq.~(\ref{boson qfi}). In the presence of pure dephasing of the probe and qubit, one can find that the QFI only deviates slightly from the Heisenberg scaling even under a significant dephasing $\gamma_P=\gamma_A=10^{-2}g$. When taking the probe relaxation into account, the Heisenberg-scaling behavior is attainable in a limited regime of $\bar{n}$, i.e., $\bar{n}\leq10$ and $\bar{n}\leq42$, under $\gamma_P=\gamma_A=\kappa=10^{-2}g$ and $10^{-3}g$, respectively. It indicates that probe relaxation is the most detrimental factor to our protocol.

\subsection{Qubit-spin-ensemble system}\label{subsection probe spin system}

In this section, we consider a metrology model consisting of a giant spin probe (spin ensemble) coupled to an ancillary spin by a general Heisenberg $XXZ$ interaction. The full Hamiltonian reads
\begin{equation} \label{qubit spin Hamiltonian}
H=\omega_PJ_z+\omega_A\sigma_z+g_zJ_z\sigma_z+g(J_+\sigma_-+J_-\sigma_+),
\end{equation}
where $J_\mu=\sum_{l=1}^N\sigma_\mu^l/2$, $\mu=x,y,z$, represents the collective spin operator with $N$ the total spin number of the probe ensemble and $\sigma_\mu^l$ the $\mu$ component of the Pauli operator for the $l$th probe spin. $\sigma_\mu$ is the Pauli matrix of the ancillary qubit. $\omega_P$ and $\omega_A$ denote the energy splitting of the probe spin and the ancillary spin, respectively.

For the spin-ensemble probe, the parametric encoding operator $R(\theta)=e^{-i\theta G}$ is typically about a rotation around a fixed axis with an angle $\theta$. Thus, it is reasonable to assume that the phase generator is a collective spin operator along a particular direction,
\begin{equation}\label{phase generator atomic system}
G=\cos\chi J_z+\sin\chi(\cos\phi J_x+\sin\phi J_y),
\end{equation}
where $\chi$ and $\phi$ are the polar and azimuthal angles, respectively. Inserting Eq.~(\ref{bigspin full evolution operator}) about the time-evolution operator for qubit-spin-ensemble system and Eq.~(\ref{phase generator atomic system}) into Eq.~(\ref{parametrization process in atomic system}) and then Eq.~(\ref{effective phase generator}), we obtain the effective dynamical generator $\mathcal{H}$, whose four components are
\begin{subequations}
\begin{align}\nonumber
&M=\cos\chi{\rm Tr}_A\left(e^{iHt}J_ze^{-iHt}\right)\\ \label{largespin matrix elements M}
+&\left[e^{-i\phi}\sin\chi{\rm Tr}_A\left(e^{iHt}J_+e^{-iHt}\right)+{\rm H.c.}\right],
\\ \nonumber
&X,Y,Z=\cos\chi{\rm Tr}_A\left(e^{iHt}J_ze^{-iHt}\sigma_{x,y,z}\right)\\ \label{largespin matrix elements XYZ}
+&\left[e^{-i\phi}\sin\chi{\rm Tr}_A\left(e^{iHt}J_+e^{-iHt}\sigma_{x,y,z}\right)+{\rm H.c.}\right].
\end{align}
\end{subequations}
Using the integer-valued polynomial~\cite{cahen1997integer}, one can find that the criterion in Eqs.~(\ref{criterion about M}) and (\ref{criterion about XYZ}), or more precisely, $M=X=Y=0$, can be deduced by $\chi=\pi/2$ and
\begin{subequations}
\begin{align}\label{criteria M in XXZ interation}
\sum_{k=0}^{k_{\rm max}}\sum_{\substack{l=0\wedge\\k-l={\rm odd}}}^{k-1}C_k^l\tilde d_km^l&=\frac{\pi}{t}\left(\sum_{k=0}^{k_{\rm max}-1}\sum_{l=0}^k\tilde a_k\frac{s(k,l)}{k!}m^l+\frac{1}{2}\right),\\\label{criteria XYZ in XXZ interation}
\sum_{k=0}^{k_{\rm max}}\sum_{\substack{l=0\wedge\\k-l={\rm even}}}^{k-2}C_k^l\tilde d_km^l&=\frac{\pi}{t}\left(\sum_{k=0}^{k_{\rm max}-2}\sum_{l=0}^k\tilde b_k\frac{s(k,l)}{k!}m^l+\frac{1}{2}\right),
\end{align}
\end{subequations}
where $\tilde a_k$ and $\tilde b_k$ are arbitrary integers, and $m=-j$, $-j+1$, $\cdots$, $j$ and $j=N/2$. The coefficient $\tilde d_k$ is the $k$th coefficient in the Taylor expansion of the function
\begin{equation}\label{spin frequency}
\begin{aligned}
\tilde\Omega(m)=&\Bigl\{(g_z^2-g^2)m^2+(g_z^2-g^2+2g_z\Delta)m\\
&+\left[\left(\frac{g_z}{2}+\Delta\right)^2+g^2j(j+1)\right]\Bigl\}^\frac{1}{2}.
\end{aligned}
\end{equation}

Following the same procedure in Sec.~\ref{subsection probe bosonic system}, one can take $k_{\rm max}=1$ and hold Eq.~(\ref{spin frequency}) up to the zeroth order in $m$, and then obtain the corresponding coefficient as
\begin{equation}\label{spin coefficients}
\begin{aligned}
\tilde\Omega(m)=&\frac{g_z^2-g^2+2g_z\Delta}{\sqrt{(g_z+2\Delta)^2+g^2N(N+2)}}m\\
&+\sqrt{\left(\frac{g_z}{2}+\Delta\right)^2+g^2j(j+1)}.
\end{aligned}
\end{equation}
In this case, Eqs.~(\ref{criteria M in XXZ interation}) and (\ref{criteria XYZ in XXZ interation}) can be deduced by the constraint equation
\begin{equation}\label{solution for criteria in XXZ interation}
\frac{t}{\pi}\frac{g_z^2-g^2+2g_z\Delta}{\sqrt{(g_z+2\Delta)^2+g^2N(N+2)}}=\tilde a_0+\frac{1}{2},
\end{equation}
for the systematic parameters.

When $M=X=Y=0$, the optimized phase generator in Eq.~(\ref{phase generator atomic system}) is found to be $G=\cos\phi J_x+\sin\phi J_y$ and the associated effective dynamical generator is formulated as
\begin{equation}\label{spin dynamical generator}
\mathcal{H}=Z\sigma_z=[\sin(\phi-\omega_Pt)J_x-\cos(\phi-\omega_Pt)J_y]\sigma_z.
\end{equation}
Consequently, QFI for the spin-ensemble probe takes the same expression as its bosonic counterpart in Eq.~(\ref{boson qfi}). The Heisenberg scaling $F_Q=N^2$ is attainable when the probe system is prepared in pure states $|j,j\rangle_Z$ and $|j,-j\rangle_Z$, or arbitrary superposed or mixed states over them, such as $(|j,j\rangle_Z+|j,-j\rangle_Z)/\sqrt{2}$ or $(|j,j\rangle_Z\langle j,j|+|j,-j\rangle_Z\langle j,-j|)/2$. Here $|j,m\rangle_Z$ with $-j\leq m\leq j$ are eigenstates of the optimized collective spin operator $Z$.

It is remarkable to find that our protocol for the qubit-spin-ensemble model is also insensitive to the deviations in the systematic parameters. In Eq.~(\ref{solution for criteria in XXZ interation}), we consider $\tilde a_0\rightarrow\tilde a'_0=\tilde a_0+\delta\tilde{a}_0$, where $|\delta\tilde{a}_0|\ll1$ indicates the deviation magnitude. Then we have
\begin{equation}
\frac{t}{\pi}\frac{g_z^2-g^2+2g_z\Delta}{\sqrt{(g_z+2\Delta)^2+g^2N(N+2)}}=\tilde a_0+\delta\tilde a_0+\frac{1}{2},
\end{equation}
and hence the frequency in Eq.~(\ref{spin coefficients}) becomes
\begin{equation}\label{Rabi frequency spin}
\tilde\Omega(m)\!\!=\!\!\left(\tilde a_0+\delta\tilde a_0+\frac{1}{2}\right)\frac{m\pi}{t}+\sqrt{\left(\frac{g_z}{2}+\Delta\right)^2+g^2j(j+1)}.
\end{equation}
Inserting Eqs.~(\ref{Rabi frequency spin}) and (\ref{bigspin full evolution operator}) into Eqs.~(\ref{largespin matrix elements M}) and (\ref{largespin matrix elements XYZ}), we obtain the four components of the effective dynamical generator share similar forms as Eqs.~(\ref{M bosonic}) and (\ref{Z bosonic}) by replacing $a/\sqrt2$ and $a^\dagger/\sqrt2$ with $J_-$ and $J_+$, respectively. Up to the leading order of $\delta\tilde a_0$, QFI for the initial spin-ensemble probe $|j,j\rangle_Z$ and the ancillary qubit $|+\rangle$ becomes
\begin{equation}
F_Q\approx N^2-N(N-1)(\pi\delta\tilde a_0)^2.
\end{equation}

\begin{figure}[htbp]
\begin{centering}
\includegraphics[width=0.8\linewidth]{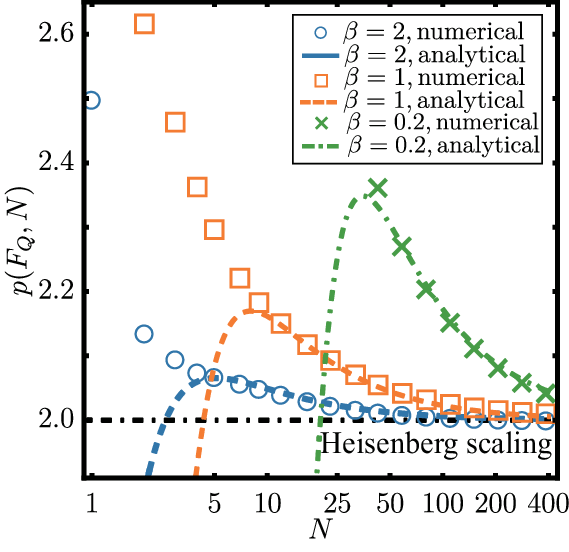}
\caption{Scaling exponent of QFI $p(F_Q,N)$ about rotation-phase estimation as a function of $N$ for a spin-ensemble in a mixed state $\rho_s^{\rm th}$. The blue circles, orange squares, and green crosses indicate the exact numerical simulations with inverse temperature $\beta=2$, $1$, and $0.2$, respectively. The blue solid line, orange dashed line, and green dot-dashed line are the relevant analytical results in Eq.~(\ref{spin ana thermal}). The black dot-dashed line indicates the Heisenberg scaling. Here $\rho_A=|+\rangle\langle+|$, $\Delta=10g$, $g_z=g$, and $t\approx\pi/(2g)$.}\label{QFI spin thermal state}
\end{centering}
\end{figure}

QFI about the estimated rotation phase can hold the Heisenberg scaling even when the probe ensemble is relaxed to be a mixed state, e.g., a thermal state in the bases of $|j,m\rangle_Z$, i.e., $\rho_s^{\rm th}=e^{-\beta Z}/\tilde{Z}_{\beta}$, where $\tilde{Z}_{\beta}={\rm Tr}[\exp(-\beta Z)]$ is the partition function. By Eqs.~(\ref{QFI for probe state and ancillary qubit}) and (\ref{spin dynamical generator}), we have
\begin{equation}\label{QFI spin thermal}
F_Q=4\frac{\sum_{m=-N/2}^{N/2}m^2\exp(-m\beta)}{\sum_{m=-N/2}^{N/2}\exp(-m\beta)}=4\frac{\partial_\beta^2S(\beta)}{S(\beta)},
\end{equation}
where $S(\beta)\equiv\sum_{m=-N/2}^{N/2}\exp(-m\beta)$ is a finite geometric series. Using the standard formula, $\sum_{m=-N/2}^{N/2}q^m=q^{-N/2}(1-q^{N+1})/(1-q)$, and setting $q=e^{-\beta}$, one can obtain a compact expression for $S(\beta)$ as well as its second derivative $\partial_\beta^2S(\beta)$. They give rise to
\begin{equation}\label{QFI define spin thermal}
\begin{aligned}
F_Q&=\frac1{(e^\beta-1)^2}\left[N^2\frac{1-e^{(3+N)\beta}}{1-e^{(1+N)\beta}}\right.+(N+2)^2\\
&\times\frac{e^{2\beta}-e^{(1+N)\beta}}{1-e^{(1+N)\beta}}
\left.+(N^2+2N-2)\frac{2e^{(2+N)\beta}-2e^\beta}{1-e^{(1+N)\beta}}\right].
\end{aligned}
\end{equation}
Then for a large-size probe, i.e., $N\gg1$, it is approximated as
\begin{equation}\label{spin ana thermal}
\begin{aligned}
F_Q&\approx\frac{e^{2\beta}N^2+(N+2)^2-2e^\beta(N^2+2N-2)}{(e^\beta-1)^2}\\
&=N^2-\frac{4}{e^\beta-1}N+4\frac{e^\beta+1}{(e^\beta-1)^2}.
\end{aligned}
\end{equation}
For $\beta\rightarrow\infty$, we have $F_Q\rightarrow N^2$; and for a finite $\beta$, we still have a quadratic dependence on $N$ at least in the low-temperature limit. The scaling behavior of QFI in Eq.~(\ref{spin ana thermal}) is confirmed by the numerical result in Fig.~\ref{QFI spin thermal state} for a sufficiently large $N$. Without loss of generality, we choose the ancillary qubit $|\varphi\rangle=|+\rangle$ and $\Delta=10g$ and $g_z=g$ in Eq.~(\ref{solution for criteria in XXZ interation}), which yield $t\approx\pi/(2g)$. It is found that, when $\beta=2$, $\beta=1$, and $\beta=0.2$, the numerical simulation can be well described by the approximate analytical results for $N\geq5$, $N\geq11$, and $N\geq40$, respectively.

\begin{figure}[htbp]
\begin{centering}
\includegraphics[width=0.8\linewidth]{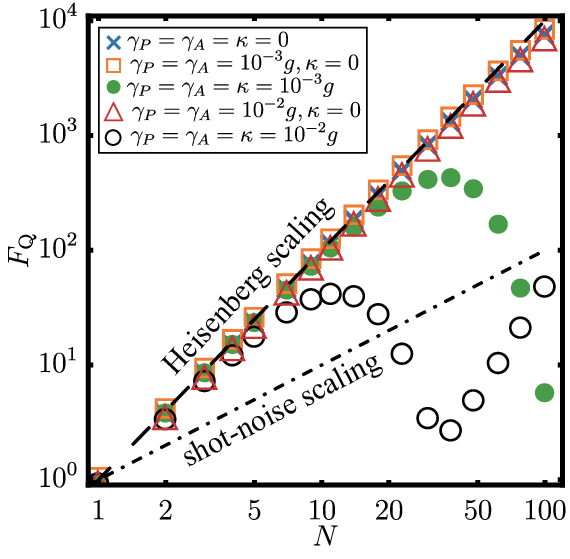}
\caption{QFI as a function of probe size $N$ for a pure state $|j,j\rangle_Z$ under various decoherence rates. The black dashed line and the black dot-dashed line indicate the Heisenberg scaling and the shot-noise scaling, respectively. The other setting is the same as Fig.~\ref{QFI spin thermal state}.}\label{QFI_dissipative_spin}
\end{centering}
\end{figure}

In the presence of decoherence, the evolution of the density matrix can be written in a master equation similar to Eq.~(\ref{masterEquation of bosonic system}) by replacing the operators $a$ and $a^\dagger a$ with $J_-$ and $J_z$, respectively, where the Hamiltonian is given by Eq.~(\ref{qubit spin Hamiltonian}). In Fig.~\ref{QFI_dissipative_spin}, QFI under various decay and dephasing rates is numerically obtained by the output state at the moment determined by Eq.~(\ref{solution for criteria in XXZ interation}). Similar to Fig.~\ref{QFI_dissipative_boson}, it is confirmed that QFI holds an asymptotic Heisenberg-scaling law of $N$ in the presence of pure dephasing of the probe and qubit. With probe dissipation, the scaling behavior of QFI gradually deviates from the Heisenberg scaling with increasing probe spin number $N$. This phenomenon can be understood by Eq.~(\ref{solution for criteria in XXZ interation}), which shows that the simulation time $t$ is approximately proportional to $N$. Consequently, decay drives the probe toward the ground state $|j,-j\rangle$ in the long-time limit.

\section{qubit-assisted metrology with encoding embedded in evolution}\label{section entangle encoding}

The deduction from Eqs.~(\ref{criterion about M}) and (\ref{criterion about XYZ}) suggests that our coupling criterion applies to a broad range of metrologies assisted by the qubit, where the joint free evolution and the parametric encoding are separable. We now consider a scenario in which the parametric encoding is embedded in the joint time evolution of the composite system, which means Eq.~(\ref{parametrization process in atomic system}) is no longer valid. Magnetometry~\cite{Jones2009magnetic,wang2015high,boto2018moving,xie2021beating,montenegro2022sequential} is a paradigmatic model about this scenario, where the probe system is subject to an unknown transverse magnetic field applied along, e.g., the $x$ direction. For example, in a Heisenberg spin chain consisting of $N$ spin-1/2 particles~\cite{montenegro2022sequential}, the unknown magnetic field is locally applied to the first spin, and its information can be extracted by performing sequential projective measurements on the last spin. And that protocol enables a sensitivity beyond SQL.

In this section, the full Hamiltonian describes a probe spin ensemble coupled to an ancillary qubit:
\begin{equation}\label{magnetometer Hamiltonian}
H=\omega_PJ_z+\theta J_x+gJ_z\sigma_z+\omega_A\sigma_z,
\end{equation}
where $\theta$ denotes the to-be-estimated magnetic-field strength. This Hamiltonian is experimentally feasible, e.g., in the nitrogen-vacancy (NV) centers~\cite{xie2021beating}, $J_z$ and $\sigma_z$ describe the $^{13}C$ nuclear spins and the NV electron spin, respectively.

In the following, our magnetometry is verified to be suitable for detecting weak magnetic fields, where the frequency shift ($\propto\theta$) induced by the target magnetic field is much smaller than the energy splitting $\omega_P$ of the probe spin. This condition is well established in geomagnetic measurement based on NV centers~\cite{yu2025experimental}, where the intrinsic probe frequency is $2.87$ GHz and the target magnetic field is around $0.421$ G associated with frequency shift of $1.18$ MHz, a thousandth of $\omega_P$. So far the sensitivity of the magnetic field reaches $10^{-6}{\rm G}/\sqrt{\rm Hz}$ based on the ensembles of NV defects~\cite{le2012efficient}. However, it remains constrained by the shot-noise limit with $1/\sqrt{N}$. Moreover, the magnetometry can also indirectly access the other quantities of interest~\cite{wang2015high,boto2018moving}, e.g., in measurement of the electric current that is inferred from the magnetic field (typically as weak as $10^0$ G) generated by a current-carrying copper wire~\cite{wang2015high}.

Using Eq.~(\ref{magnetometer Hamiltonian}), the composite time-evolution operator can be expressed in the eigenbasis of $\sigma_z$, i.e.,
\begin{equation}\label{evolution magnetometer}
\begin{aligned}
U_\theta(t)=&e^{-i\omega_At\sigma_z}\big(e^{-i\theta_+J_y}e^{-i\omega_+tJ_z}e^{i\theta_+J_y}|e\rangle\langle e|\\
&+e^{-i\theta_-J_y}e^{-i\omega_-tJ_z}e^{i\theta_-J_y}|g\rangle\langle g|\big),
\end{aligned}
\end{equation}
where
\begin{equation}\label{coefficient for evolution}
\theta_\pm\equiv\arccos\left(\frac{\omega_P\pm g}{\omega_\pm}\right),\quad
\omega_\pm\equiv\sqrt{(\omega_P\pm g)^2+\theta^2}.
\end{equation}
Inserting Eq.~(\ref{evolution magnetometer}) into Eq.~(\ref{effective phase generator}), one can obtain the effective dynamical generator $\mathcal{H}$, in which the four components read $X=Y=0$ and
\begin{subequations}
\begin{align}\label{magnetometer component M}
M &=\frac{X_++X_-}{2}J_x+\frac{Y_++Y_-}{2}J_y+\frac{Z_++Z_-}{2}J_z,\\\label{magnetometer component Z}
Z &=\frac{X_+-X_-}{2}J_x+\frac{Y_+-Y_-}{2}J_y+\frac{Z_+-Z_-}{2}J_z,
\end{align}
\end{subequations}
with
\begin{equation}\label{coefficient for matrix}
\begin{aligned}
X_\pm &\equiv t\partial_\theta\omega_\pm\sin\theta_\pm+\partial_\theta\theta_{\pm}\sin(\omega_\pm t)\cos\theta_\pm,\\
Y_\pm &\equiv \partial_\theta\theta_\pm[\cos(\omega_{\pm}t)-1] ,\\
Z_\pm &\equiv t\partial_\theta\omega_\pm\cos\theta_\pm-\partial_{\theta}\theta_{\pm}\sin(\omega_\pm t)\sin\theta_\pm.
\end{aligned}
\end{equation}
One can find that they have met the criterion in Eq.~(\ref{criterion about XYZ}).

In the low magnetic-field regime, i.e., $\theta\ll\omega_P$, Eq.~(\ref{coefficient for matrix}) becomes
\begin{equation}\label{approximate coefficient for matrix}
\begin{aligned}
X_\pm&\approx{\rm sign}(\theta)\frac{\sin[t(\omega_P\pm g) ]}{\omega_P\pm g},\\
Y_\pm&\approx\pm{\rm sign}(\theta)\frac{\cos[t(\omega_P\pm g)]-1}{\omega_P\pm g},\\
Z_\pm&\approx0,
\end{aligned}
\end{equation}
up to the zeroth order of $\theta$, where ${\rm sign}(\theta)$ indicates the sign of $\theta$. Inserting Eqs.~(\ref{magnetometer component M}) and (\ref{approximate coefficient for matrix}) into Eq.~(\ref{criterion about M}), we have
\begin{equation}
\begin{aligned}
g\cos(\omega_P t)\sin(gt)-\omega_P\sin(\omega_P t)\cos(gt)&=0,\\
g[\cos(\omega_P t)\cos(gt)-1]+\omega_P\sin(\omega_P t)\sin(gt)&=0,
\end{aligned}
\end{equation}
and their solutions are
\begin{equation}\label{solution for magnetometer}
\omega_Pt=(n+2n_P)\pi,\quad gt=(n+2n_g)\pi,
\end{equation}
where $n_P$ and $n_g$ are two independent integers. Using Eq.~(\ref{solution for magnetometer}), however, one can find that Eq.~(\ref{magnetometer component Z}) reduces to $Z\approx 0$, which implies $\mathcal{H}=Z\sigma_z\approx 0$. To obtain a nontrivial effective dynamical generator, Eq.~(\ref{solution for magnetometer}) can be slightly modified to be
\begin{equation}\label{approximate solution for magnetometer}
\omega_Pt=n\pi+2n_P\pi+\delta_P,\quad gt=n\pi+2n_g\pi+\delta_g,
\end{equation}
where $\delta_P$ and $\delta_g$ are dimensionless coefficients. To the first order in $\delta_P$ and $\delta_g$, the effective dynamical generator can be written as
\begin{equation}
\mathcal{H}=M+Z\sigma_z\approx{\rm sign}(\theta)\frac{\delta_g t}{(n+2n_P)\pi}J_x\sigma_z,
\end{equation}
which is constrained by
\begin{equation}\label{ratio between probe frequency and couple}
\frac{\delta_P}{\delta_g}=\frac{n+2n_g}{n+2n_P}.
\end{equation}
Consequently, QFI in Eq.~(\ref{QFI for probe state and ancillary qubit}) becomes
\begin{equation}\label{magnetometer qfi}
F_Q\approx4A{\rm Tr}\left(J_x^2\rho_P\right),
\end{equation}
with $A=[\delta_g t/(n+2n_P)\pi]^2$.

For a small deviation from the dimensionless ratio $b_0\equiv\delta_P/\delta_g$, denoted by $\delta b_0$ with $|\delta b_0/b_0|\ll1$, the constraint condition in Eq.~(\ref{ratio between probe frequency and couple}) becomes
\begin{equation}\label{ratio deviation}
\frac{\delta_P}{\delta_g}=b_0+\delta b_0=\frac{n+2n_g}{n+2n_P}+\delta b_0.
\end{equation}
Substituting Eqs.~(\ref{approximate solution for magnetometer}) and (\ref{ratio deviation}) into Eq.~(\ref{approximate coefficient for matrix}), one can obtain the coefficients in the low magnetic-field regime,
\begin{equation}\label{coefficients XYZ}
\begin{aligned}
X_\pm&\approx\pm\frac{{\rm sign}(\theta)\delta_gt}{(n+2n_P)\pi}\left[1\pm\frac{n+2n_P}{(n+2n_P)\pm(n+2n_g)}\delta b_0\right],\\
Y_\pm&\approx Z_\pm\approx0,
\end{aligned}
\end{equation}
up to the first order of $\delta_g$ and the second order of $\delta b_0$. Using Eq.~(\ref{magnetometer component M}), (\ref{magnetometer component Z}), and (\ref{coefficients XYZ}), one can find that the components are $X=Y=0$ and
\begin{equation}
\begin{aligned}
M&\approx{\rm sign}(\theta)\frac{\delta_gt}{(n+2n_P)\pi}J_x\frac{(n+2n_P)^2}{4(n+n_P+n_g)(n_P-n_g)}\delta b_0,\\
Z&\approx{\rm sign}(\theta)\frac{\delta_gt}{(n+2n_P)\pi}J_x(1-C_n\delta b_0),
\end{aligned}
\end{equation}
with the coefficient $C_n=(n+2n_P)(n+2n_g)/[4(n+n_P+n_g)(n_P-n_g)]$. Consequently, with the initial spin-ensemble probe $|j,j\rangle_x$ and the ancillary qubit $|+\rangle$, the QFI reads
\begin{equation}\label{qfi magnetometry}
F_Q\approx AN^2(1-C_n\delta b_0)^2.
\end{equation}
In practice, the probe frequency $\omega_P$ is much larger than the qubit-probe coupling strength $g$. Then we have $n_P\gg n, n_g$ due to the constraint for $\omega_P$ and $g$ in Eq.~(\ref{approximate solution for magnetometer}). Consequently, we have $C_n\ll1$, which indicates that the Heisenberg scaling behavior in Eq.~(\ref{qfi magnetometry}) is insensitive to the deviation of the ratio $\delta_P/\delta_g$.

\begin{figure}[htbp]
\begin{centering}
\includegraphics[width=0.8\linewidth]{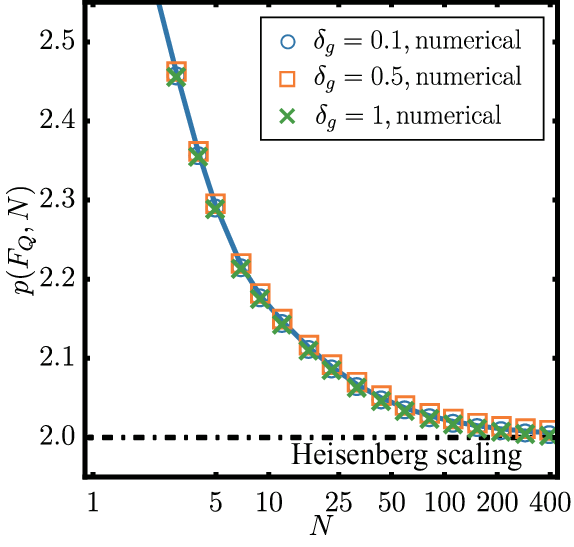}
\caption{Scaling exponent of QFI $p(F_Q,N)$ for magnetometry as a function of $N$ for a spin-ensemble prepared in a mixed state, e.g., the thermal state $\rho_m^{\rm th}$ with $\beta=1$. The blue circles, orange squares, and green crosses indicate the exact numerical simulations with systematic parameter $\delta_g=0.1$, $0.5$, and $1$, respectively. The blue solid line is the approximate result in Eq.~(\ref{magnetometer qfi thermal}). The black dot-dashed line indicates the Heisenberg scaling. Here $\rho_A=|+\rangle\langle+|$, $t=\pi/2$, and $\omega_P\approx10g$.}\label{QFI magnetometer thermal state}
\end{centering}
\end{figure}

If the probe system is initially in a mixed state, e.g., the thermal state $\rho_m^{\rm th}=e^{-\beta J_x}/Z'_{\beta}$ with $Z'_{\beta}={\rm Tr}[\exp(-\beta J_x)]$ the partition function, then, through a calculation similar to that in Eq.~(\ref{QFI define spin thermal}), the QFI in Eq.~(\ref{magnetometer qfi}) for a thermal probe state $\rho_m^{\rm th}$ is
\begin{equation}\label{magnetometer qfi thermal}
\begin{aligned}
F_Q&=\frac{A}{(e^\beta-1)^2}\Bigg[N^2\frac{1-e^{(3+N)\beta}}{1-e^{(1+N)\beta}}\\
&+(N+2)^2\frac{e^{2\beta}-e^{(1+N)\beta}}{1-e^{(1+N)\beta}}\\
&+(N^2+2N-2)\frac{2e^{(2+N)\beta}-2e^\beta}{1-e^{(1+N)\beta}}\Bigg].
\end{aligned}
\end{equation}
Then for a large-size probe, i.e., $N\gg1$, it reduces to Eq.~(\ref{spin ana thermal}) in addition to an extra prefactor $A$. The asymptotic square scaling law of $N$ exhibited in Eq.~(\ref{magnetometer qfi thermal}) can be justified by the numerical simulation in Fig.~\ref{QFI magnetometer thermal state}, which shows the scaling exponent of the QFI as a function of probe size $N$ under various $\delta_g$'s and a fixed $\beta=1$. We choose $t=\pi/2$, $n=0$, $n_P=10$, and $n_g=1$ in Eq.~(\ref{approximate solution for magnetometer}), which yield $\omega_P\approx10g$. One cannot distinguish the numerical results under various $\delta_g$'s. It is consistent with Eq.~(\ref{magnetometer qfi}) that the systematic parameter $\delta_g$, or more precisely, $\omega_P$ and $g$ merely affect the prefactor of the QFI, while the scaling exponent $p(F_Q,N)$ approaches the Heisenberg limit for a sufficiently large $N$.

\begin{figure}[htbp]
\begin{centering}
\includegraphics[width=0.8\linewidth]{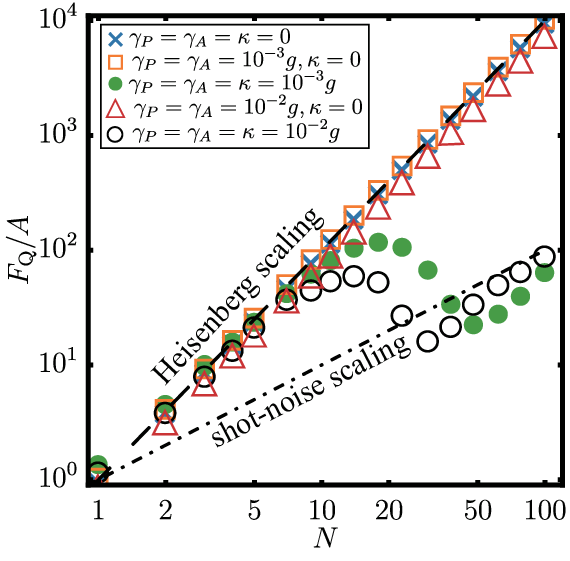}
\caption{Renormalized QFI $F_Q/A$ as a function of probe size $N$ for a pure state $|j,j\rangle_x$ under decoherence with various decay and dephasing rates. The black dashed line and the black dot-dashed line indicate the Heisenberg scaling and the shot-noise scaling, respectively. The other setting is the same as Fig.~\ref{QFI magnetometer thermal state}.}\label{QFI_dissipative_magnetometry}
\end{centering}
\end{figure}

To investigate the influence of the external decoherence on our magnetometry (similar to Sec.~\ref{subsection probe spin system}), we numerically calculate the QFI of the output state by Lindblad master equation with the Hamiltonian given by Eq.~(\ref{magnetometer Hamiltonian}). In Fig.~\ref{QFI_dissipative_magnetometry}, we take $\delta_g=0.1$ and present the renormalized QFI $F_Q/A$ for the small parameter $\theta\approx0$ under various decay and dephasing rates, where the probe is initialized as $|j,j\rangle_x$. The scaling behavior of the QFI is qualitatively similar to those in Fig.~\ref{QFI_dissipative_spin}. In the presence of the probe relaxation, the Heisenberg scaling still dominates for $N\leq6$ and $N\leq12$ under $\gamma_P=\gamma_A=\kappa=10^{-2}g$ and $10^{-3}g$, respectively. With increasing probe spin number $N$, the behavior of the QFI gradually drops to the shot-noise scaling. In particular, we have $F_Q\approx N$ for $N = 22$ and $N=38$, for $\gamma_P=\gamma_A=\kappa=10^{-2}g$ and $10^{-3}g$, respectively.

\section{Discussion and Conclusion}\label{section conclusion}

In summary, we have established a general coupling criterion for quantum metrology assisted by a two-level system interacting with a probe system, either the bosonic-mode probe or spin-ensemble probe, with arbitrarily initial state, by which the quantum Fisher information for the estimated parameters encoded in the probe system can be optimized to be the mean square of a parameter dependent dynamical generator with respect to the initial state of the composite system. Our criterion is confirmed by two scenarios: (i) the parametric encoding and the composite time evolution are performed separately, and (ii) the parametric encoding cannot be separated from the composite evolution.

In scenario (i), by employing the integer-valued polynomials, we find that the probe system prepared in a mixed state, e.g., a thermal state, can be used to attain an asymptotic Heisenberg scaling behavior in parametric estimation with respect to the relevant setting, namely the average photon number $\bar{n}$ for bosonic systems or the probe size $N$ for spin ensembles. In scenario (ii), we propose a qubit-assisted magnetometry scheme. For the weak magnetic fields, Heisenberg scaling can be achieved by using a spin ensemble in mixed states, a proper state of the ancillary qubit, and optimized parameters. Our work suggests that metrologically relevant resource states can be dynamically generated through the qubit-probe interaction and the estimation precision can be pushed to exceed the standard quantum limit by an almost classical input state in the qubit-assisted paradigms. And our protocol is insensitive to both dephasing and imperfections.

The precision ensured by our criterion can be attained via a practical measurement. In Appendix~\ref{appendix classical Fisher information}, we provide a time-reversal strategy by which the classical Fisher information in our metrology protocol can attain the Heisenberg scaling and saturate its quantum counterpart upon a projective measurement on the ancillary qubit or on the composite system.

\section*{Acknowledgments}

We acknowledge grant support from the National Natural Science Foundation of China (Grant No. U25A20199) and the ``Pioneer'' and ``Leading Goose'' R\&D of Zhejiang Province (Grant No. 2025C01028).

\appendix
\section{Time evolution operator for qubit-assisted systems}\label{appendix evolution operator}

This appendix contributes to deriving the time-evolution operators for qubit-assisted hybrid systems associated with the general JC interaction and Heisenberg $XXZ$ interaction in Eqs.~(\ref{qubit boson hamiltionian}) and (\ref{qubit spin Hamiltonian}), respectively.

We first recall Eq.~(\ref{qubit boson hamiltionian}), that could be decomposed as the free Hamiltonian and interaction Hamiltonian,
\begin{equation}\label{qubit boson hamiltionian appendix}
\begin{aligned}
&H=H_0+H_I,\\
&H_0=\omega_Pa^\dagger a+\omega_A\sigma_z,\\
&H_I=g_za^\dagger a\sigma_z+g\left(a^\dagger\sigma_-+a\sigma_+\right).
\end{aligned}
\end{equation}
Due to the commutation relation $[H, a^\dagger a+\sigma_z/2]=0$, the composite time evolution operator can be partitioned into the subspaces $|0,g\rangle\oplus\{|n+1,g\rangle,|n,e\rangle\}$ with $n=0$, $1$, $2$, $\cdots$. The Hamiltonian in a typical subspace $\{|n+1,g\rangle,|n,e\rangle\}$ can be written as
\begin{equation}\label{subspace in qubit boson}
h_n=A_n+\left(
\begin{array}{cc}
-\Lambda_n&\omega_n\\
\omega_n&\Lambda_n
\end{array}
\right) ,
\end{equation}
with $A_n=(n+1/2)\omega_P-g_z/2$, $\omega_n=g\sqrt{n+1}$, $\Lambda_n=(n+1/2)g_z+\Delta$, and $\Delta\equiv\omega_A-\omega_P/2$. The local evolution operator in the relevant subspace thus reads
\begin{equation}
\begin{aligned}
&e^{-ih_nt}=e^{-iA_nt}\\
&\times\left(
\begin{array}{cc}
\cos(\Omega_nt)+\frac{i\Lambda_n}{\Omega_n}\sin(\Omega_nt)&-\frac{i\omega_n}{\Omega_n}\sin(\Omega_nt)\\
-\frac{i\omega_n}{\Omega_n}\sin(\Omega_nt)&\cos(\Omega_nt)-\frac{i\Lambda_n}{\Omega_n}\sin(\Omega_nt)
\end{array}
\right),
\end{aligned}
\end{equation}
with $\Omega_n=\sqrt{\Lambda_n^2+\omega_n^2}$.

Consequently, the time evolution operator in the whole Hilbert space can be organized as
\begin{equation}
e^{-iHt}=\sum_{n=0}^\infty e^{-ih_nt}+e^{i\omega_At}|0,g\rangle\langle0,g|.
\end{equation}
Using $\omega(\hat{n})\equiv g\sqrt{\hat{n}+1}$, $\Lambda(\hat{n})\equiv g_z(\hat{n}+1/2)+\Delta$, $\Omega(\hat{n})\equiv\sqrt{\omega^2(\hat{n})+\Lambda^2(\hat{n})}$, and $A(\hat{n})\equiv\omega_P(\hat{n}+1/2)-g_z/2$ about the number operator $\hat{n}\equiv a^\dagger a$, the composite time-evolution operator can be rewritten in a more explicit form,
\begin{equation}\label{expression full evolution operator}
\begin{aligned}
&e^{-iHt}=\Bigg\{\left[\cos\Omega(\hat{n})t-\frac{i\Lambda(\hat{n})}{\Omega(\hat{n})}\sin\Omega(\hat{n})t\right]
\otimes|e\rangle\langle e|\\
-&ie^{i\omega_Pt}g\frac{\sin\Omega(\hat{n})t}{\Omega(\hat{n})}a\otimes|e\rangle\langle g|
-iga^\dagger\frac{\sin\Omega(\hat{n})t}{\Omega(\hat{n})}\otimes|g\rangle\langle e|\\
+&\left[\cos\Omega(\hat{n}-1)t+\frac{i\Lambda(\hat{n}-1)}{\Omega(\hat{n}-1)}\sin\Omega(\hat{n}-1)t\right]\\
&\otimes|g\rangle\langle g|e^{i\omega_Pt}\Bigg\}e^{-iA(\hat{n})t}.
\end{aligned}
\end{equation}

Similarly, the composite time evolution operator about the Heisenberg XXZ interaction given in Eq.~(\ref{qubit spin Hamiltonian}) can be partitioned into the subspaces $|j,j,e\rangle\oplus |j,-j,g\rangle\oplus\{|j,m+1,g\rangle ,|j,m,e\rangle\}$ with $-j\leq m\leq j-1$ for the quantum number $j=N/2$. $N$ is the size of the probe spin ensemble. In a typical subspace $\{|j,m+1,g\rangle ,|j,m,e\rangle\}$, the full Hamiltonian can be written in a form similar to Eq.~(\ref{subspace in qubit boson}) by replacing $n$ with $m$ and replacing $\omega_n$ with $\tilde{\omega}_m=g\sqrt{(j-m)(j+m+1)}$.

Consequently, the time evolution operator in the whole Hilbert space can be expressed in terms of the functions of $J_z$, i.e.,
\begin{equation}\label{bigspin full evolution operator}
\begin{aligned}
e^{-iHt}&=\Bigg\{\left[\cos\tilde{\Omega}(J_z)t-\frac{i\Lambda(J_z)}{\tilde{\Omega}(J_z)}
\sin\tilde{\Omega}(J_z)t\right]\otimes|e\rangle\langle e|\\
&-ie^{i\omega_Pt}g\frac{\sin\tilde{\Omega}(J_z)t}{\tilde{\Omega}(J_z)}J_-\otimes|e\rangle\langle g|\\
&-igJ_+\frac{\sin\tilde{\Omega}(J_z)t}{\tilde{\Omega}(J_z)}\otimes|g\rangle\langle e|\\
&+\left[\cos\tilde{\Omega}(J_z-1)t+\frac{i\Lambda(J_z-1)}{\tilde{\Omega}(J_z-1)}\sin\tilde{\Omega}(J_z-1)t\right]\\
&\otimes|g\rangle\langle g|e^{i\omega_Pt}\Bigg\}e^{-iA(J_z)t},
\end{aligned}
\end{equation}
where
\begin{equation}\label{function of operator}
\begin{aligned}
\tilde{\Omega}(J_z)&\equiv\sqrt{\tilde{\omega}^2(J_z)+\Lambda^2(J_z)},\\
\tilde{\omega}(J_z)&\equiv g\sqrt{(j-J_z)(j+J_z+1)},
\end{aligned}
\end{equation}
and $\Lambda(J_z)$ and $A(J_z)$ are defined similar to $\Lambda(\hat{n})$ and $A(\hat{n})$, respectively.

\section{Explicit criterion in qubit-assisted systems}\label{appendix solution for criteria}

This appendix contributes to deriving the explicit expression of the coupling criterion for the general qubit-oscillator system and qubit-spin-ensemble system given by Eqs.~(\ref{qubit boson hamiltionian}) and (\ref{qubit spin Hamiltonian}), respectively. The criterion consists of two conditions, i.e., $M=0$ in Eq.~(\ref{criterion about M}) and $X$, $Y$, or $Z=0$ in Eq.~(\ref{criterion about XYZ}).

We first consider the criterion in Eq.~(\ref{criterion about M}). Inserting Eq.~(\ref{expression full evolution operator}) into Eq.~(\ref{boson matrix elements M}), one can enforce that
\begin{equation}\label{boson condition M}
\begin{aligned}
&e^{i\Omega(a^\dagger a+1)t}+e^{i\Omega(a^\dagger a-1)t}\\
=&\sum_{n=0}^\infty\left(e^{i\Omega(n+1)t}+e^{i\Omega(n-1)t}\right)|n\rangle\langle n|=0,
\end{aligned}
\end{equation}
where we used the relation $a^\dagger a=\sum_{n=0}^\infty n|n\rangle\langle n|$ and the operator function $\Omega(\hat{n})=\Omega(a^\dagger a)$ in Eq.~(\ref{boson frequency_definition}). Using the Taylor expansion, the function $\Omega(n)$ in Eq.~(\ref{boson condition M}) can be rewritten as
\begin{equation}\label{Omega n}
\begin{aligned}
&\Omega(n)=\sqrt{\omega^2(n)+\Lambda^2(n)}\\
=&\sqrt{g_z^2n^2+(g^2+g_z^2+2g_z\Delta)n+\left[g^2+\left(\frac{g_z}{2}+\Delta\right)^2\right]}\\
=&\sum_{k=0}^\infty d_kn^k,
\end{aligned}
\end{equation}
where the coefficient $d_k$ is the $k$th coefficient. Substituting Eq.~(\ref{Omega n}) to Eq.~(\ref{boson condition M}), one can find that Eq.~(\ref{criterion about M}), i.e., the criterion $M=0$, is equivalent to
\begin{equation}
\left[\sum_{k=0}^\infty\left(\sum_{\substack{l=0\wedge\\k-l={\rm odd}}}^kC_k^ld_kn^l\right)\right]t=\pi\left(u+\frac{1}{2}\right),
\end{equation}
where $C_k^l=k!/[l!(k-l)!]$ denotes the binomial coefficient and $u$ is an arbitrary integer. According to the theory of integer-valued polynomials~\cite{cahen1997integer}, the integer $u$ can be presented as a polynomial expansion in the integer variable $n$ and consequently we have
\begin{equation}\label{condition M expression}
\sum_{k=0}^\infty\left(\sum_{\substack{l=0\wedge\\k-l={\rm odd}}}^kC_k^ld_kn^l\right)=\frac{\pi}{t}\left(\sum_{k=0}^\infty\sum_{l=0}^ka_k\frac{s(k,l)}{k!}n^l+\frac{1}{2}\right),
\end{equation}
where $a_k$ is an arbitrary integer and $s(k,l)$ indicates the Stirling numbers of the first kind.

Similarly, by the time-evolution operator in Eq.~(\ref{expression full evolution operator}) and Eq.~(\ref{boson matrix elements XYZ}), the three components of the effective dynamical generator in Eq.~(\ref{criterion about XYZ}) read
\begin{equation}\label{boson expression X}
\begin{aligned}
&X=\frac{e^{-i\xi}{\rm Tr}_A (e^{iHt}ae^{-iHt}\sigma_x)+{\rm H.c.}}{2\sqrt2}\\
=&\frac{ige^{-i(\xi +\omega_Pt)}}{4\sqrt2}\left[\frac{1}{\Omega(n)}+\frac{1}{\Omega(n+1)}\right]\\
\times&{\rm Im}\left(e^{i\Omega(n)t}e^{-i\Omega(n+1)t}\right)a^2+{\rm H.c.}\\
+&\frac{g\left[e^{-i(\xi+\omega_Pt)}+e^{i(\xi+\omega_Pt)}\right]}{4\sqrt2}
\Bigl\{\frac{\Delta}{\Omega(n-1)\Omega(n)}\\
\times&\left({\rm Re}\left[e^{i\Omega(n-1)t}e^{-i\Omega(n)t}\right]-{\rm Re}\left[e^{i\Omega(n-1)t}e^{i\Omega(n)t}\right]\right)\\
+&i\left[\frac{n}{\Omega(n-1)}+\frac{n+1}{\Omega(n)}\right]{\rm Im}\left[e^{i\Omega(n-1)t}e^{-i\Omega(n)t}\right]\Bigl\},
\end{aligned}
\end{equation}
\begin{equation}\label{boson expression Y}
\begin{aligned}
&Y=\frac{e^{-i\xi}{\rm Tr}_A (e^{iHt}ae^{-iHt}\sigma_y)+{\rm H.c.}}{2\sqrt2}\\
=&\frac{g}{4\sqrt2}e^{-i(\xi+\omega_Pt)}\left[\frac{1}{\Omega(n)}+\frac{1}{\Omega(n+1)}\right]\\
\times&{\rm Im}\left(e^{i\Omega(n)t}e^{-i\Omega(n+1)t}\right)a^2+{\rm H.c.} \\
+&\frac{g\left[e^{-i(\xi+\omega_Pt)}-e^{i(\xi+\omega_Pt)}\right]}{4\sqrt2}\Big\{\frac{i\Delta}{\Omega(n-1)\Omega(n)}\\
\times&\left({\rm Re}\left[e^{i\Omega(n-1)t}e^{-i\Omega(n)t}\right]-{\rm Re}\left[e^{i\Omega(n-1)t}e^{i\Omega(n)t}\right]\right)\\
-&\left[\frac{n}{\Omega(n-1)}+\frac{n+1}{\Omega(n)}\right]{\rm Im}\left[e^{i\Omega(n-1)t}e^{-i\Omega(n)t}\right]\Big\},
\end{aligned}
\end{equation}
and
\begin{equation}\label{boson expression Z}
\begin{aligned}
&Z=\frac{e^{-i\xi}{\rm Tr}_A (e^{iHt}ae^{-iHt}\sigma_z)+{\rm H.c.}}{2\sqrt2}\\
=&\frac{e^{-i(\xi+\omega_Pt)}}{4\sqrt2}\Big
\{\left[2+\frac{g^2(n+2)+\Delta^2}{\Omega(n)\Omega(n+1)}
+\frac{g^2n+\Delta^2}{\Omega(n-1)\Omega(n)}\right]\\
\times&{\rm Re}\left(e^{i\Omega(n)t}e^{-i\Omega(n+1)t}\right)\\
-&i\left[\frac{\Delta}{\Omega(n-1)}+2\frac{\Delta}{\Omega(n)}
+\frac{\Delta}{\Omega(n+1)}\right]\\
\times&{\rm Im}\left[e^{i\Omega(n)t}e^{-i\Omega(n+1)t}\right]\Big\}a+{\rm H.c.}
\end{aligned}
\end{equation}
It is found that $Z\neq0$ and $X=0$ or $Y=0$ can be deduced by
\begin{equation}\label{appendix XY=0}
{\rm Im}\left(e^{i\Omega(n)t}e^{-i\Omega(n+1)t}\right)=0.
\end{equation}
Through a similar analysis below Eq.~(\ref{boson condition M}), we have
\begin{equation}\label{condition XY expression}
\sum_{k=0}^\infty\sum_{\substack{l=0\wedge\\k-l={\rm even}}}^{k-2}C_k^ld_kn^l=\frac{\pi}{t}\left(\sum_{k=0}^\infty\sum_{l=0}^kb_k\frac{s(k,l)}{k!}n^l+\frac{1}{2}\right),
\end{equation}
where $b_k$ is an arbitrary integer and $k$ ranges from one to infinite as an integer. For the computational convenience, one can set a finite cutoff $k_{\rm max}$. Consequently, Eqs.~(\ref{condition M expression}) and (\ref{condition XY expression}) reduce to Eqs.~(\ref{solution for criteria boson M}) and (\ref{solution for criteria boson XY}), respectively.

As for the qubit-spin-ensemble system in Eq.~(\ref{qubit spin Hamiltonian}), one can insert the time-evolution operator in Eq.~(\ref{bigspin full evolution operator}) into Eq.~(\ref{largespin matrix elements M}) and then obtain the expression for the component $M$ of the effective dynamical generator. It is found that the criterion in Eq.~(\ref{criterion about M}), i.e., $M=0$, can be deduced by $\chi=\pi/2$ and a form similar to Eq.~(\ref{boson condition M}) by replacing $n$ with $m$ and replacing $\Omega(a^\dagger a\pm1)$ with $\tilde\Omega(J_z\pm1)$ in Eq.~(\ref{spin frequency}):
\begin{equation}\label{largespin condition M}
\begin{aligned}
&e^{i\tilde\Omega(J_z+1)t}+e^{i\tilde\Omega(J_z-1)t}\\
=&\sum_{m=-j}^j\left(e^{i\tilde\Omega(m+1)t}+e^{i\tilde\Omega(m-1)t}\right)|j,m\rangle\langle j,m|=0,
\end{aligned}
\end{equation}
where we have used $J_z=\sum_{m=-j}^jm|j,m\rangle\langle j,m|$. According to the theory of integer-valued polynomials~\cite{cahen1997integer} and the Taylor expansion of $\tilde\Omega(m)$ with respect to $m$, Eq.~(\ref{largespin condition M}) can be reduced to
\begin{equation}
\sum_{k=0}^\infty\sum_{\substack{l=0\wedge\\k-l={\rm odd}}}^{k-1}C_k^l\tilde d_km^l=\frac{\pi}{t}\left(\sum_{k=0}^\infty\sum_{l=0}^k\tilde a_k\frac{s(k,l)}{k!}m^l+\frac{1}{2}\right),
\end{equation}
where $\tilde d_k$ indicates the $k$th coefficient. With a finite cutoff $k_{\rm max}$, it becomes Eq.~(\ref{criteria M in XXZ interation}) in the main text.

Using Eqs.~(\ref{bigspin full evolution operator}) and (\ref{largespin matrix elements XYZ}), one can obtain the three components $X$, $Y$, and $Z$ of the effective dynamical generator, similar to Eqs.~(\ref{boson expression X}), (\ref{boson expression Y}), and (\ref{boson expression Z}), respectively, by replacing $\xi$ with $\phi$, $\Omega(n)$ with $\tilde\Omega(m)$, and $a(a^\dagger)$ with $J_-(J_+)$. We find that $Z\neq0$ and $X=0$ or $Y=0$ can be deduced by
\begin{equation}
{\rm Im}\left(e^{i\tilde\Omega(m)t}e^{-i\tilde\Omega(m+1)t}\right)=0.
\end{equation}
Using the integer-valued polynomials~\cite{cahen1997integer} and the Taylor expansion of $\tilde\Omega(m)$ with a finite cutoff $k_{\rm max}$, one can obtain Eq.~(\ref{criteria XYZ in XXZ interation}) in the main text.

\section{Classical Fisher information for the qubit-spin-ensemble system }\label{appendix classical Fisher information}

This appendix contributes to a practical measurement scheme that can extract all the parameter information encoded in the composite system. Here we consider the qubit-spin-ensemble system studied in Sec.~\ref{subsection probe spin system} and employ the time-reversal strategy~\cite{yurke19862,pezze2017optimal,chen2024qubit} by introducing a second stage of joint evolution $U(t_2)$ after the parametric encoding. The whole derivation can be directly extended to the other two models in the main text.

Upon the joint evolution $U(t_2)$ and the phase generator $G$ in Eq.~(\ref{phase generator atomic system}) with $\chi=\pi/2$ and $\phi=0$, the entire evolution operator in Eq.~(\ref{parametrization process in atomic system}) becomes
\begin{equation}\label{entire evolution}
U_\theta=U(t_2)R(\theta)U(t)=e^{-iHt_2}e^{-i\theta J_x}e^{-iH t}.
\end{equation}
It has been shown~\cite{yurke19862,pezze2017optimal,chen2024qubit} that the CFI can saturate the QFI when the joint evolution operator $U(t_2)$ becomes the time reversal of $U(t)$, i.e., $U(t_2)=U^\dagger(t)$ or $U(t)U(t_2)=e^{-iH(t+t_2)}=\mathcal I^{2(N+1)}$ up to a global phase. With the optimized duration $t$ in Eq.~(\ref{solution for criteria in XXZ interation}), it renders $t_2=(4n_2-1)t$ with $n_2$ integer. Then the entire evolution operator in Eq.~(\ref{entire evolution}) becomes
\begin{equation}\label{optimized entire evolution}
U_\theta=\exp(-i\theta Z\sigma_z),
\end{equation}
where the collective angular momentum operator $Z$ is given in Eq.~(\ref{spin dynamical generator}).

When the probe system is in the eigenstate $|j,j\rangle_Z$ of the operator $Z$ and the ancillary qubit is prepared at the pure state $|+\rangle$, the output state after the entire evolution~(\ref{optimized entire evolution}) reads
\begin{equation}
\begin{aligned}
&\rho(\theta)=U_\theta\rho_P\otimes\rho_AU_\theta^\dagger\\
=&\frac{1}{2}|j,j\rangle_Z\langle j,j|\otimes\left(e^{-i\theta j}|e\rangle +e^{i\theta j}|g\rangle\right)\left(e^{i\theta j}\langle e|+e^{-i\theta j}\langle g|\right).
\end{aligned}
\end{equation}
Performing the projective measurements $|\pm\rangle\langle\pm|$ on the qubit, the probability of detecting the qubit in $|\pm\rangle$ reads
\begin{equation}
P(\pm|\theta)=\langle\pm|\rho(\theta)|\pm\rangle=\frac{1\pm\cos(N\theta)}{2},
\end{equation}
by which CFI saturates its quantum counterpart as
\begin{equation}
F_C=\frac{[dP(+|\theta)/d\theta]^2}{P(+|\theta)}+\frac{[dP(-|\theta)/d\theta]^2}{P(-|\theta)}=N^2=F_Q.
\end{equation}
It indicates that only performing the projective measurement on the ancillary qubit can extract all information about the estimated parameter.

If the spin-ensemble probe is initialized as a finite-temperature thermal state, i.e., $\rho^{\rm th}_s$, then the output state reads
\begin{equation}
\begin{aligned}
\rho(\theta)=&U_\theta\rho^{\rm th}_s\otimes\rho_AU_\theta^\dagger\\
&=\sum_{m=-j}^j\frac{e^{-m\beta}}{2\tilde{Z}_\beta}|j,m\rangle_Z\langle j,m|\\
\otimes&\left(e^{-i\theta m}|e\rangle +e^{i\theta m}|g\rangle\right)\left(e^{i\theta m}\langle e|+e^{-i\theta m}\langle g|\right).
\end{aligned}
\end{equation}
Here we perform the projective measurements $|\Psi_{m,\pm}\rangle\langle\Psi_{m,\pm}|$ on the full system with $|\Psi_{m,\pm}\rangle=|j,m\rangle_Z\otimes|\pm\rangle$. The probability of detecting the output state as  $|\Psi_{m,\pm}\rangle$ reads
\begin{equation}
P(m,\pm|\theta)=\langle\Psi_{m,\pm}|\rho(\theta)|\Psi_{m,\pm}\rangle=e^{-m\beta}\frac{1\pm\cos(2m\theta)}{2\tilde{Z}_\beta}.
\end{equation}
Consequently, one can obtain the CFI as
\begin{equation}
\begin{aligned}
F_C=&2\sum_{m=-j}^j\frac{m^2e^{-m\beta}}{\tilde{Z}_\beta}\sin^2(2m\theta)\\
\times&\left[\frac{1}{1+\cos(2m\theta)}+\frac{1}{1-\cos(2m\theta)}\right]\\
=&4\sum_{m=-j}^j\frac{m^2e^{-m\beta}}{\tilde{Z}_\beta}=F_Q.
\end{aligned}
\end{equation}
The last line is exactly the same as Eq.~(\ref{QFI spin thermal}).

\bibliographystyle{apsrevlong}
\bibliography{ref}

\begin{thebibliography}{56}%
\makeatletter
\providecommand \@ifxundefined [1]{%
 \@ifx{#1\undefined}
}%
\providecommand \@ifnum [1]{%
 \ifnum #1\expandafter \@firstoftwo
 \else \expandafter \@secondoftwo
 \fi
}%
\providecommand \@ifx [1]{%
 \ifx #1\expandafter \@firstoftwo
 \else \expandafter \@secondoftwo
 \fi
}%
\providecommand \natexlab [1]{#1}%
\providecommand \enquote  [1]{``#1''}%
\providecommand \bibnamefont  [1]{#1}%
\providecommand \bibfnamefont [1]{#1}%
\providecommand \citenamefont [1]{#1}%
\providecommand \href@noop [0]{\@secondoftwo}%
\providecommand \href [0]{\begingroup \@sanitize@url \@href}%
\providecommand \@href[1]{\@@startlink{#1}\@@href}%
\providecommand \@@href[1]{\endgroup#1\@@endlink}%
\providecommand \@sanitize@url [0]{\catcode `\\12\catcode `\$12\catcode
  `\&12\catcode `\#12\catcode `\^12\catcode `\_12\catcode `\%12\relax}%
\providecommand \@@startlink[1]{}%
\providecommand \@@endlink[0]{}%
\providecommand \url  [0]{\begingroup\@sanitize@url \@url }%
\providecommand \@url [1]{\endgroup\@href {#1}{\urlprefix }}%
\providecommand \urlprefix  [0]{URL }%
\providecommand \Eprint [0]{\href }%
\providecommand \doibase [0]{http://dx.doi.org/}%
\providecommand \selectlanguage [0]{\@gobble}%
\providecommand \bibinfo  [0]{\@secondoftwo}%
\providecommand \bibfield  [0]{\@secondoftwo}%
\providecommand \translation [1]{[#1]}%
\providecommand \BibitemOpen [0]{}%
\providecommand \bibitemStop [0]{}%
\providecommand \bibitemNoStop [0]{.\EOS\space}%
\providecommand \EOS [0]{\spacefactor3000\relax}%
\providecommand \BibitemShut  [1]{\csname bibitem#1\endcsname}%
\let\auto@bib@innerbib\@empty
\bibitem [{\citenamefont {Giovannetti}\ \emph {et~al.}(2006)\citenamefont
  {Giovannetti}, \citenamefont {Lloyd},\ and\ \citenamefont
  {Maccone}}]{giovannetti2006quantum}%
  \BibitemOpen
  \bibfield  {author} {\bibinfo {author} {\bibfnamefont {V.}~\bibnamefont
  {Giovannetti}}, \bibinfo {author} {\bibfnamefont {S.}~\bibnamefont {Lloyd}},
  \ and\ \bibinfo {author} {\bibfnamefont {L.}~\bibnamefont {Maccone}},\
  }\bibfield  {title} {\emph {\bibinfo {title} {Quantum metrology},\ }}\href
  {\doibase 10.1103/PhysRevLett.96.010401} {\bibfield  {journal} {\bibinfo
  {journal} {Phys. Rev. Lett.}\ }\textbf {\bibinfo {volume} {96}},\ \bibinfo
  {pages} {010401} (\bibinfo {year} {2006})}\BibitemShut {NoStop}%
\bibitem [{\citenamefont {Budker}\ and\ \citenamefont
  {Romalis}(2007)}]{budker2007optical}%
  \BibitemOpen
  \bibfield  {author} {\bibinfo {author} {\bibfnamefont {D.}~\bibnamefont
  {Budker}}\ and\ \bibinfo {author} {\bibfnamefont {M.}~\bibnamefont
  {Romalis}},\ }\bibfield  {title} {\emph {\bibinfo {title} {Optical
  magnetometry},\ }}\href {\doibase 10.1038/nphys566} {\bibfield  {journal}
  {\bibinfo  {journal} {Nat. phys.}\ }\textbf {\bibinfo {volume} {3}},\
  \bibinfo {pages} {227} (\bibinfo {year} {2007})}\BibitemShut {NoStop}%
\bibitem [{\citenamefont {Paris}(2009)}]{paris2009quantum}%
  \BibitemOpen
  \bibfield  {author} {\bibinfo {author} {\bibfnamefont {M.~G.}\ \bibnamefont
  {Paris}},\ }\bibfield  {title} {\emph {\bibinfo {title} {Quantum estimation
  for quantum technology},\ }}\href {\doibase 10.1142/S0219749909004839}
  {\bibfield  {journal} {\bibinfo  {journal} {Int. J. Quantum Inf.}\ }\textbf
  {\bibinfo {volume} {7}},\ \bibinfo {pages} {125} (\bibinfo {year}
  {2009})}\BibitemShut {NoStop}%
\bibitem [{\citenamefont {Degen}\ \emph {et~al.}(2017)\citenamefont {Degen},
  \citenamefont {Reinhard},\ and\ \citenamefont
  {Cappellaro}}]{degen2017quantum}%
  \BibitemOpen
  \bibfield  {author} {\bibinfo {author} {\bibfnamefont {C.~L.}\ \bibnamefont
  {Degen}}, \bibinfo {author} {\bibfnamefont {F.}~\bibnamefont {Reinhard}}, \
  and\ \bibinfo {author} {\bibfnamefont {P.}~\bibnamefont {Cappellaro}},\
  }\bibfield  {title} {\emph {\bibinfo {title} {Quantum sensing},\ }}\href
  {\doibase 10.1103/RevModPhys.89.035002} {\bibfield  {journal} {\bibinfo
  {journal} {Rev. Mod. Phys.}\ }\textbf {\bibinfo {volume} {89}},\ \bibinfo
  {pages} {035002} (\bibinfo {year} {2017})}\BibitemShut {NoStop}%
\bibitem [{\citenamefont {Katori}(2011)}]{Katori2011optical}%
  \BibitemOpen
  \bibfield  {author} {\bibinfo {author} {\bibfnamefont {H.}~\bibnamefont
  {Katori}},\ }\bibfield  {title} {\emph {\bibinfo {title} {Optical lattice
  clocks and quantum metrology},\ }}\href {\doibase 10.1038/nphoton.2011.45}
  {\bibfield  {journal} {\bibinfo  {journal} {Nat. Photon.}\ }\textbf {\bibinfo
  {volume} {5}},\ \bibinfo {pages} {203} (\bibinfo {year} {2011})}\BibitemShut
  {NoStop}%
\bibitem [{\citenamefont {Ludlow}\ \emph {et~al.}(2015)\citenamefont {Ludlow},
  \citenamefont {Boyd}, \citenamefont {Ye}, \citenamefont {Peik},\ and\
  \citenamefont {Schmidt}}]{Ludlow2015optical}%
  \BibitemOpen
  \bibfield  {author} {\bibinfo {author} {\bibfnamefont {A.~D.}\ \bibnamefont
  {Ludlow}}, \bibinfo {author} {\bibfnamefont {M.~M.}\ \bibnamefont {Boyd}},
  \bibinfo {author} {\bibfnamefont {J.}~\bibnamefont {Ye}}, \bibinfo {author}
  {\bibfnamefont {E.}~\bibnamefont {Peik}}, \ and\ \bibinfo {author}
  {\bibfnamefont {P.~O.}\ \bibnamefont {Schmidt}},\ }\bibfield  {title} {\emph
  {\bibinfo {title} {Optical atomic clocks},\ }}\href {\doibase
  10.1103/RevModPhys.87.637} {\bibfield  {journal} {\bibinfo  {journal} {Rev.
  Mod. Phys.}\ }\textbf {\bibinfo {volume} {87}},\ \bibinfo {pages} {637}
  (\bibinfo {year} {2015})}\BibitemShut {NoStop}%
\bibitem [{\citenamefont {Caves}(1981)}]{Caves1981quantum}%
  \BibitemOpen
  \bibfield  {author} {\bibinfo {author} {\bibfnamefont {C.~M.}\ \bibnamefont
  {Caves}},\ }\bibfield  {title} {\emph {\bibinfo {title} {Quantum-mechanical
  noise in an interferometer},\ }}\href {\doibase 10.1103/PhysRevD.23.1693}
  {\bibfield  {journal} {\bibinfo  {journal} {Phys. Rev. D}\ }\textbf {\bibinfo
  {volume} {23}},\ \bibinfo {pages} {1693} (\bibinfo {year}
  {1981})}\BibitemShut {NoStop}%
\bibitem [{\citenamefont {Taylor}\ and\ \citenamefont
  {Bowen}(2016)}]{Taylor2016quantum}%
  \BibitemOpen
  \bibfield  {author} {\bibinfo {author} {\bibfnamefont {M.~A.}\ \bibnamefont
  {Taylor}}\ and\ \bibinfo {author} {\bibfnamefont {W.~P.}\ \bibnamefont
  {Bowen}},\ }\bibfield  {title} {\emph {\bibinfo {title} {Quantum metrology
  and its application in biology},\ }}\href {\doibase
  https://doi.org/10.1016/j.physrep.2015.12.002} {\bibfield  {journal}
  {\bibinfo  {journal} {Phys. Rep.}\ }\textbf {\bibinfo {volume} {615}},\
  \bibinfo {pages} {1} (\bibinfo {year} {2016})}\BibitemShut {NoStop}%
\bibitem [{\citenamefont {Mauranyapin}\ \emph {et~al.}(2017)\citenamefont
  {Mauranyapin}, \citenamefont {Madsen}, \citenamefont {Taylor}, \citenamefont
  {Waleed},\ and\ \citenamefont {Bowen}}]{Mauranyapin2017evanescent}%
  \BibitemOpen
  \bibfield  {author} {\bibinfo {author} {\bibfnamefont {N.}~\bibnamefont
  {Mauranyapin}}, \bibinfo {author} {\bibfnamefont {L.}~\bibnamefont {Madsen}},
  \bibinfo {author} {\bibfnamefont {M.}~\bibnamefont {Taylor}}, \bibinfo
  {author} {\bibfnamefont {M.}~\bibnamefont {Waleed}}, \ and\ \bibinfo {author}
  {\bibfnamefont {W.}~\bibnamefont {Bowen}},\ }\bibfield  {title} {\emph
  {\bibinfo {title} {Evanescent single-molecule biosensing with quantum-limited
  precision},\ }}\href {\doibase 10.1038/nphoton.2017.99} {\bibfield  {journal}
  {\bibinfo  {journal} {Nat. Photon.}\ }\textbf {\bibinfo {volume} {11}},\
  \bibinfo {pages} {477} (\bibinfo {year} {2017})}\BibitemShut {NoStop}%
\bibitem [{\citenamefont {Jones}\ \emph {et~al.}(2009)\citenamefont {Jones},
  \citenamefont {Karlen}, \citenamefont {Fitzsimons}, \citenamefont {Ardavan},
  \citenamefont {Benjamin}, \citenamefont {Briggs},\ and\ \citenamefont
  {Morton}}]{Jones2009magnetic}%
  \BibitemOpen
  \bibfield  {author} {\bibinfo {author} {\bibfnamefont {J.~A.}\ \bibnamefont
  {Jones}}, \bibinfo {author} {\bibfnamefont {S.~D.}\ \bibnamefont {Karlen}},
  \bibinfo {author} {\bibfnamefont {J.}~\bibnamefont {Fitzsimons}}, \bibinfo
  {author} {\bibfnamefont {A.}~\bibnamefont {Ardavan}}, \bibinfo {author}
  {\bibfnamefont {S.~C.}\ \bibnamefont {Benjamin}}, \bibinfo {author}
  {\bibfnamefont {G.~A.~D.}\ \bibnamefont {Briggs}}, \ and\ \bibinfo {author}
  {\bibfnamefont {J.~J.}\ \bibnamefont {Morton}},\ }\bibfield  {title} {\emph
  {\bibinfo {title} {Magnetic field sensing beyond the standard quantum limit
  using 10-spin noon states},\ }}\href {\doibase 10.1126/science.1170730}
  {\bibfield  {journal} {\bibinfo  {journal} {Science}\ }\textbf {\bibinfo
  {volume} {324}},\ \bibinfo {pages} {1166} (\bibinfo {year}
  {2009})}\BibitemShut {NoStop}%
\bibitem [{\citenamefont {Giovannetti}\ \emph {et~al.}(2004)\citenamefont
  {Giovannetti}, \citenamefont {Lloyd},\ and\ \citenamefont
  {Maccone}}]{giovannetti2004quantum}%
  \BibitemOpen
  \bibfield  {author} {\bibinfo {author} {\bibfnamefont {V.}~\bibnamefont
  {Giovannetti}}, \bibinfo {author} {\bibfnamefont {S.}~\bibnamefont {Lloyd}},
  \ and\ \bibinfo {author} {\bibfnamefont {L.}~\bibnamefont {Maccone}},\
  }\bibfield  {title} {\emph {\bibinfo {title} {Quantum-enhanced measurements:
  beating the standard quantum limit},\ }}\href {\doibase
  10.1126/science.1104149} {\bibfield  {journal} {\bibinfo  {journal}
  {Science}\ }\textbf {\bibinfo {volume} {306}},\ \bibinfo {pages} {1330}
  (\bibinfo {year} {2004})}\BibitemShut {NoStop}%
\bibitem [{\citenamefont {T{\'o}th}\ and\ \citenamefont
  {Apellaniz}(2014)}]{toth2014quantum}%
  \BibitemOpen
  \bibfield  {author} {\bibinfo {author} {\bibfnamefont {G.}~\bibnamefont
  {T{\'o}th}}\ and\ \bibinfo {author} {\bibfnamefont {I.}~\bibnamefont
  {Apellaniz}},\ }\bibfield  {title} {\emph {\bibinfo {title} {Quantum
  metrology from a quantum information science perspective},\ }}\href {\doibase
  10.1088/1751-8113/47/42/424006} {\bibfield  {journal} {\bibinfo  {journal}
  {J. Phys. A: Math. Theor.}\ }\textbf {\bibinfo {volume} {47}},\ \bibinfo
  {pages} {424006} (\bibinfo {year} {2014})}\BibitemShut {NoStop}%
\bibitem [{\citenamefont {Nawrocki}(2019)}]{nawrocki2015introduction}%
  \BibitemOpen
  \bibfield  {author} {\bibinfo {author} {\bibfnamefont {W.}~\bibnamefont
  {Nawrocki}},\ }\href@noop {} {\emph {\bibinfo {title} {{Introduction to
  Quantum Metrology}}}},\ \bibinfo {edition} {2nd}\ ed.\ (\bibinfo  {publisher}
  {Springer Nature Switzerland},\ \bibinfo {address} {Cham, Switzerland},\
  \bibinfo {year} {2019})\BibitemShut {NoStop}%
\bibitem [{\citenamefont {Giovannetti}\ \emph {et~al.}(2011)\citenamefont
  {Giovannetti}, \citenamefont {Lloyd},\ and\ \citenamefont
  {Maccone}}]{giovannetti2011advances}%
  \BibitemOpen
  \bibfield  {author} {\bibinfo {author} {\bibfnamefont {V.}~\bibnamefont
  {Giovannetti}}, \bibinfo {author} {\bibfnamefont {S.}~\bibnamefont {Lloyd}},
  \ and\ \bibinfo {author} {\bibfnamefont {L.}~\bibnamefont {Maccone}},\
  }\bibfield  {title} {\emph {\bibinfo {title} {Advances in quantum
  metrology},\ }}\href {\doibase 10.1038/nphoton.2011.35} {\bibfield  {journal}
  {\bibinfo  {journal} {Nat. photon.}\ }\textbf {\bibinfo {volume} {5}},\
  \bibinfo {pages} {222} (\bibinfo {year} {2011})}\BibitemShut {NoStop}%
\bibitem [{\citenamefont {Pezz\`e}\ \emph {et~al.}(2018)\citenamefont
  {Pezz\`e}, \citenamefont {Smerzi}, \citenamefont {Oberthaler}, \citenamefont
  {Schmied},\ and\ \citenamefont {Treutlein}}]{pezze2018quantum}%
  \BibitemOpen
  \bibfield  {author} {\bibinfo {author} {\bibfnamefont {L.}~\bibnamefont
  {Pezz\`e}}, \bibinfo {author} {\bibfnamefont {A.}~\bibnamefont {Smerzi}},
  \bibinfo {author} {\bibfnamefont {M.~K.}\ \bibnamefont {Oberthaler}},
  \bibinfo {author} {\bibfnamefont {R.}~\bibnamefont {Schmied}}, \ and\
  \bibinfo {author} {\bibfnamefont {P.}~\bibnamefont {Treutlein}},\ }\bibfield
  {title} {\emph {\bibinfo {title} {Quantum metrology with nonclassical states
  of atomic ensembles},\ }}\href {\doibase 10.1103/RevModPhys.90.035005}
  {\bibfield  {journal} {\bibinfo  {journal} {Rev. Mod. Phys.}\ }\textbf
  {\bibinfo {volume} {90}},\ \bibinfo {pages} {035005} (\bibinfo {year}
  {2018})}\BibitemShut {NoStop}%
\bibitem [{\citenamefont {Polino}\ \emph {et~al.}(2020)\citenamefont {Polino},
  \citenamefont {Valeri}, \citenamefont {Spagnolo},\ and\ \citenamefont
  {Sciarrino}}]{polino2020photonic}%
  \BibitemOpen
  \bibfield  {author} {\bibinfo {author} {\bibfnamefont {E.}~\bibnamefont
  {Polino}}, \bibinfo {author} {\bibfnamefont {M.}~\bibnamefont {Valeri}},
  \bibinfo {author} {\bibfnamefont {N.}~\bibnamefont {Spagnolo}}, \ and\
  \bibinfo {author} {\bibfnamefont {F.}~\bibnamefont {Sciarrino}},\ }\bibfield
  {title} {\emph {\bibinfo {title} {Photonic quantum metrology},\ }}\href
  {https://doi.org/10.1116/5.0007577} {\bibfield  {journal} {\bibinfo
  {journal} {AVS Quantum Sci.}\ }\textbf {\bibinfo {volume} {2}},\ \bibinfo
  {pages} {024703} (\bibinfo {year} {2020})}\BibitemShut {NoStop}%
\bibitem [{\citenamefont {Ye}\ and\ \citenamefont
  {Zoller}(2024)}]{ye2024essay}%
  \BibitemOpen
  \bibfield  {author} {\bibinfo {author} {\bibfnamefont {J.}~\bibnamefont
  {Ye}}\ and\ \bibinfo {author} {\bibfnamefont {P.}~\bibnamefont {Zoller}},\
  }\bibfield  {title} {\emph {\bibinfo {title} {Essay: Quantum sensing with
  atomic, molecular, and optical platforms for fundamental physics},\ }}\href
  {\doibase 10.1103/PhysRevLett.132.190001} {\bibfield  {journal} {\bibinfo
  {journal} {Phys. Rev. Lett.}\ }\textbf {\bibinfo {volume} {132}},\ \bibinfo
  {pages} {190001} (\bibinfo {year} {2024})}\BibitemShut {NoStop}%
\bibitem [{\citenamefont {DeMille}\ \emph {et~al.}(2024)\citenamefont
  {DeMille}, \citenamefont {Hutzler}, \citenamefont {Rey},\ and\ \citenamefont
  {Zelevinsky}}]{demille2024quantum}%
  \BibitemOpen
  \bibfield  {author} {\bibinfo {author} {\bibfnamefont {D.}~\bibnamefont
  {DeMille}}, \bibinfo {author} {\bibfnamefont {N.~R.}\ \bibnamefont
  {Hutzler}}, \bibinfo {author} {\bibfnamefont {A.~M.}\ \bibnamefont {Rey}}, \
  and\ \bibinfo {author} {\bibfnamefont {T.}~\bibnamefont {Zelevinsky}},\
  }\bibfield  {title} {\emph {\bibinfo {title} {Quantum sensing and metrology
  for fundamental physics with molecules},\ }}\href {\doibase
  10.1038/s41567-024-02499-9} {\bibfield  {journal} {\bibinfo  {journal} {Nat.
  Phys.}\ }\textbf {\bibinfo {volume} {20}},\ \bibinfo {pages} {741} (\bibinfo
  {year} {2024})}\BibitemShut {NoStop}%
\bibitem [{\citenamefont {Genoni}\ \emph {et~al.}(2013)\citenamefont {Genoni},
  \citenamefont {Paris}, \citenamefont {Adesso}, \citenamefont {Nha},
  \citenamefont {Knight},\ and\ \citenamefont {Kim}}]{genoni2013optimal}%
  \BibitemOpen
  \bibfield  {author} {\bibinfo {author} {\bibfnamefont {M.~G.}\ \bibnamefont
  {Genoni}}, \bibinfo {author} {\bibfnamefont {M.~G.}\ \bibnamefont {Paris}},
  \bibinfo {author} {\bibfnamefont {G.}~\bibnamefont {Adesso}}, \bibinfo
  {author} {\bibfnamefont {H.}~\bibnamefont {Nha}}, \bibinfo {author}
  {\bibfnamefont {P.~L.}\ \bibnamefont {Knight}}, \ and\ \bibinfo {author}
  {\bibfnamefont {M.}~\bibnamefont {Kim}},\ }\bibfield  {title} {\emph
  {\bibinfo {title} {Optimal estimation of joint parameters in phase space},\
  }}\href {\doibase 10.1103/PhysRevA.87.012107} {\bibfield  {journal} {\bibinfo
   {journal} {Phys. Rev. A}\ }\textbf {\bibinfo {volume} {87}},\ \bibinfo
  {pages} {012107} (\bibinfo {year} {2013})}\BibitemShut {NoStop}%
\bibitem [{\citenamefont {Duivenvoorden}\ \emph {et~al.}(2017)\citenamefont
  {Duivenvoorden}, \citenamefont {Terhal},\ and\ \citenamefont
  {Weigand}}]{duivenvoorden2017single}%
  \BibitemOpen
  \bibfield  {author} {\bibinfo {author} {\bibfnamefont {K.}~\bibnamefont
  {Duivenvoorden}}, \bibinfo {author} {\bibfnamefont {B.~M.}\ \bibnamefont
  {Terhal}}, \ and\ \bibinfo {author} {\bibfnamefont {D.}~\bibnamefont
  {Weigand}},\ }\bibfield  {title} {\emph {\bibinfo {title} {Single-mode
  displacement sensor},\ }}\href {\doibase 10.1103/PhysRevA.95.012305}
  {\bibfield  {journal} {\bibinfo  {journal} {Phys. Rev. A}\ }\textbf {\bibinfo
  {volume} {95}},\ \bibinfo {pages} {012305} (\bibinfo {year}
  {2017})}\BibitemShut {NoStop}%
\bibitem [{\citenamefont {Bond}\ \emph {et~al.}(2026)\citenamefont {Bond},
  \citenamefont {Valahu}, \citenamefont {Shankar}, \citenamefont {Tan},\ and\
  \citenamefont {Safavi-Naini}}]{bond2025optimal}%
  \BibitemOpen
  \bibfield  {author} {\bibinfo {author} {\bibfnamefont {L.~J.}\ \bibnamefont
  {Bond}}, \bibinfo {author} {\bibfnamefont {C.~H.}\ \bibnamefont {Valahu}},
  \bibinfo {author} {\bibfnamefont {A.}~\bibnamefont {Shankar}}, \bibinfo
  {author} {\bibfnamefont {T.~R.}\ \bibnamefont {Tan}}, \ and\ \bibinfo
  {author} {\bibfnamefont {A.}~\bibnamefont {Safavi-Naini}},\ }\bibfield
  {title} {\emph {\bibinfo {title} {Optimal displacement sensing with
  spin-dependent squeezed states},\ }}\href {\doibase 10.1103/mhvv-pr5d}
  {\bibfield  {journal} {\bibinfo  {journal} {PRX Quantum}\ }\textbf {\bibinfo
  {volume} {7}},\ \bibinfo {pages} {020368} (\bibinfo {year}
  {2026})}\BibitemShut {NoStop}%
\bibitem [{\citenamefont {Wolf}\ \emph {et~al.}(2019)\citenamefont {Wolf},
  \citenamefont {Shi}, \citenamefont {Heip}, \citenamefont {Gessner},
  \citenamefont {Pezz{\`e}}, \citenamefont {Smerzi}, \citenamefont {Schulte},
  \citenamefont {Hammerer},\ and\ \citenamefont {Schmidt}}]{wolf2019motional}%
  \BibitemOpen
  \bibfield  {author} {\bibinfo {author} {\bibfnamefont {F.}~\bibnamefont
  {Wolf}}, \bibinfo {author} {\bibfnamefont {C.}~\bibnamefont {Shi}}, \bibinfo
  {author} {\bibfnamefont {J.~C.}\ \bibnamefont {Heip}}, \bibinfo {author}
  {\bibfnamefont {M.}~\bibnamefont {Gessner}}, \bibinfo {author} {\bibfnamefont
  {L.}~\bibnamefont {Pezz{\`e}}}, \bibinfo {author} {\bibfnamefont
  {A.}~\bibnamefont {Smerzi}}, \bibinfo {author} {\bibfnamefont
  {M.}~\bibnamefont {Schulte}}, \bibinfo {author} {\bibfnamefont
  {K.}~\bibnamefont {Hammerer}}, \ and\ \bibinfo {author} {\bibfnamefont
  {P.~O.}\ \bibnamefont {Schmidt}},\ }\bibfield  {title} {\emph {\bibinfo
  {title} {Motional fock states for quantum-enhanced amplitude and phase
  measurements with trapped ions},\ }}\href {\doibase
  10.1038/s41467-019-10576-4} {\bibfield  {journal} {\bibinfo  {journal} {Nat.
  commun.}\ }\textbf {\bibinfo {volume} {10}},\ \bibinfo {pages} {2929}
  (\bibinfo {year} {2019})}\BibitemShut {NoStop}%
\bibitem [{\citenamefont {Valahu}\ \emph {et~al.}(2025)\citenamefont {Valahu},
  \citenamefont {Stafford}, \citenamefont {Huang}, \citenamefont {Matsos},
  \citenamefont {Millican}, \citenamefont {Chalermpusitarak}, \citenamefont
  {Menicucci}, \citenamefont {Combes}, \citenamefont {Baragiola},\ and\
  \citenamefont {Tan}}]{valahu2025quantum}%
  \BibitemOpen
  \bibfield  {author} {\bibinfo {author} {\bibfnamefont {C.~H.}\ \bibnamefont
  {Valahu}}, \bibinfo {author} {\bibfnamefont {M.~P.}\ \bibnamefont
  {Stafford}}, \bibinfo {author} {\bibfnamefont {Z.}~\bibnamefont {Huang}},
  \bibinfo {author} {\bibfnamefont {V.~G.}\ \bibnamefont {Matsos}}, \bibinfo
  {author} {\bibfnamefont {M.~J.}\ \bibnamefont {Millican}}, \bibinfo {author}
  {\bibfnamefont {T.}~\bibnamefont {Chalermpusitarak}}, \bibinfo {author}
  {\bibfnamefont {N.~C.}\ \bibnamefont {Menicucci}}, \bibinfo {author}
  {\bibfnamefont {J.}~\bibnamefont {Combes}}, \bibinfo {author} {\bibfnamefont
  {B.~Q.}\ \bibnamefont {Baragiola}}, \ and\ \bibinfo {author} {\bibfnamefont
  {T.~R.}\ \bibnamefont {Tan}},\ }\bibfield  {title} {\emph {\bibinfo {title}
  {Quantum-enhanced multiparameter sensing in a single mode},\ }}\href
  {\doibase 10.1126/sciadv.adw9757} {\bibfield  {journal} {\bibinfo  {journal}
  {Sci. Adv.}\ }\textbf {\bibinfo {volume} {11}},\ \bibinfo {pages} {eadw9757}
  (\bibinfo {year} {2025})}\BibitemShut {NoStop}%
\bibitem [{\citenamefont {Higgins}\ \emph {et~al.}(2007)\citenamefont
  {Higgins}, \citenamefont {Berry}, \citenamefont {Bartlett}, \citenamefont
  {Wiseman},\ and\ \citenamefont {Pryde}}]{higgins2007entanglement}%
  \BibitemOpen
  \bibfield  {author} {\bibinfo {author} {\bibfnamefont {B.~L.}\ \bibnamefont
  {Higgins}}, \bibinfo {author} {\bibfnamefont {D.~W.}\ \bibnamefont {Berry}},
  \bibinfo {author} {\bibfnamefont {S.~D.}\ \bibnamefont {Bartlett}}, \bibinfo
  {author} {\bibfnamefont {H.~M.}\ \bibnamefont {Wiseman}}, \ and\ \bibinfo
  {author} {\bibfnamefont {G.~J.}\ \bibnamefont {Pryde}},\ }\bibfield  {title}
  {\emph {\bibinfo {title} {Entanglement-free heisenberg-limited phase
  estimation},\ }}\href {\doibase 10.1038/nature06257} {\bibfield  {journal}
  {\bibinfo  {journal} {Nature}\ }\textbf {\bibinfo {volume} {450}},\ \bibinfo
  {pages} {393} (\bibinfo {year} {2007})}\BibitemShut {NoStop}%
\bibitem [{\citenamefont {Resch}\ \emph {et~al.}(2007)\citenamefont {Resch},
  \citenamefont {Pregnell}, \citenamefont {Prevedel}, \citenamefont
  {Gilchrist}, \citenamefont {Pryde}, \citenamefont {O'Brien},\ and\
  \citenamefont {White}}]{resch2007time}%
  \BibitemOpen
  \bibfield  {author} {\bibinfo {author} {\bibfnamefont {K.~J.}\ \bibnamefont
  {Resch}}, \bibinfo {author} {\bibfnamefont {K.~L.}\ \bibnamefont {Pregnell}},
  \bibinfo {author} {\bibfnamefont {R.}~\bibnamefont {Prevedel}}, \bibinfo
  {author} {\bibfnamefont {A.}~\bibnamefont {Gilchrist}}, \bibinfo {author}
  {\bibfnamefont {G.~J.}\ \bibnamefont {Pryde}}, \bibinfo {author}
  {\bibfnamefont {J.~L.}\ \bibnamefont {O'Brien}}, \ and\ \bibinfo {author}
  {\bibfnamefont {A.~G.}\ \bibnamefont {White}},\ }\bibfield  {title} {\emph
  {\bibinfo {title} {Time-reversal and super-resolving phase measurements},\
  }}\href {\doibase 10.1103/PhysRevLett.98.223601} {\bibfield  {journal}
  {\bibinfo  {journal} {Phys. Rev. Lett.}\ }\textbf {\bibinfo {volume} {98}},\
  \bibinfo {pages} {223601} (\bibinfo {year} {2007})}\BibitemShut {NoStop}%
\bibitem [{\citenamefont {Xu}\ \emph {et~al.}(2013)\citenamefont {Xu},
  \citenamefont {Kedem}, \citenamefont {Sun}, \citenamefont {Vaidman},
  \citenamefont {Li},\ and\ \citenamefont {Guo}}]{xu2013phase}%
  \BibitemOpen
  \bibfield  {author} {\bibinfo {author} {\bibfnamefont {X.-Y.}\ \bibnamefont
  {Xu}}, \bibinfo {author} {\bibfnamefont {Y.}~\bibnamefont {Kedem}}, \bibinfo
  {author} {\bibfnamefont {K.}~\bibnamefont {Sun}}, \bibinfo {author}
  {\bibfnamefont {L.}~\bibnamefont {Vaidman}}, \bibinfo {author} {\bibfnamefont
  {C.-F.}\ \bibnamefont {Li}}, \ and\ \bibinfo {author} {\bibfnamefont {G.-C.}\
  \bibnamefont {Guo}},\ }\bibfield  {title} {\emph {\bibinfo {title} {Phase
  estimation with weak measurement using a white light source},\ }}\href
  {\doibase 10.1103/PhysRevLett.111.033604} {\bibfield  {journal} {\bibinfo
  {journal} {Phys. Rev. Lett.}\ }\textbf {\bibinfo {volume} {111}},\ \bibinfo
  {pages} {033604} (\bibinfo {year} {2013})}\BibitemShut {NoStop}%
\bibitem [{\citenamefont {Gross}\ \emph {et~al.}(2010)\citenamefont {Gross},
  \citenamefont {Zibold}, \citenamefont {Nicklas}, \citenamefont {Esteve},\
  and\ \citenamefont {Oberthaler}}]{gross2010nonlinear}%
  \BibitemOpen
  \bibfield  {author} {\bibinfo {author} {\bibfnamefont {C.}~\bibnamefont
  {Gross}}, \bibinfo {author} {\bibfnamefont {T.}~\bibnamefont {Zibold}},
  \bibinfo {author} {\bibfnamefont {E.}~\bibnamefont {Nicklas}}, \bibinfo
  {author} {\bibfnamefont {J.}~\bibnamefont {Esteve}}, \ and\ \bibinfo {author}
  {\bibfnamefont {M.~K.}\ \bibnamefont {Oberthaler}},\ }\bibfield  {title}
  {\emph {\bibinfo {title} {Nonlinear atom interferometer surpasses classical
  precision limit},\ }}\href {\doibase 10.1038/nature08919} {\bibfield
  {journal} {\bibinfo  {journal} {Nature}\ }\textbf {\bibinfo {volume} {464}},\
  \bibinfo {pages} {1165} (\bibinfo {year} {2010})}\BibitemShut {NoStop}%
\bibitem [{\citenamefont {d’Armagnac~de Castanet}\ \emph
  {et~al.}(2024)\citenamefont {d’Armagnac~de Castanet}, \citenamefont
  {Des~Cognets}, \citenamefont {Arguel}, \citenamefont {Templier},
  \citenamefont {Jarlaud}, \citenamefont {M{\'e}noret}, \citenamefont
  {Desruelle}, \citenamefont {Bouyer},\ and\ \citenamefont
  {Battelier}}]{d2024atom}%
  \BibitemOpen
  \bibfield  {author} {\bibinfo {author} {\bibfnamefont {Q.}~\bibnamefont
  {d’Armagnac~de Castanet}}, \bibinfo {author} {\bibfnamefont
  {C.}~\bibnamefont {Des~Cognets}}, \bibinfo {author} {\bibfnamefont
  {R.}~\bibnamefont {Arguel}}, \bibinfo {author} {\bibfnamefont
  {S.}~\bibnamefont {Templier}}, \bibinfo {author} {\bibfnamefont
  {V.}~\bibnamefont {Jarlaud}}, \bibinfo {author} {\bibfnamefont
  {V.}~\bibnamefont {M{\'e}noret}}, \bibinfo {author} {\bibfnamefont
  {B.}~\bibnamefont {Desruelle}}, \bibinfo {author} {\bibfnamefont
  {P.}~\bibnamefont {Bouyer}}, \ and\ \bibinfo {author} {\bibfnamefont
  {B.}~\bibnamefont {Battelier}},\ }\bibfield  {title} {\emph {\bibinfo {title}
  {Atom interferometry at arbitrary orientations and rotation rates},\ }}\href
  {\doibase 10.1038/s41467-024-50804-0} {\bibfield  {journal} {\bibinfo
  {journal} {Nat. Commun.}\ }\textbf {\bibinfo {volume} {15}},\ \bibinfo
  {pages} {6406} (\bibinfo {year} {2024})}\BibitemShut {NoStop}%
\bibitem [{\citenamefont {Shaw}\ \emph {et~al.}(2024)\citenamefont {Shaw},
  \citenamefont {Finkelstein}, \citenamefont {Tsai}, \citenamefont {Scholl},
  \citenamefont {Yoon}, \citenamefont {Choi},\ and\ \citenamefont
  {Endres}}]{shaw2024multi}%
  \BibitemOpen
  \bibfield  {author} {\bibinfo {author} {\bibfnamefont {A.~L.}\ \bibnamefont
  {Shaw}}, \bibinfo {author} {\bibfnamefont {R.}~\bibnamefont {Finkelstein}},
  \bibinfo {author} {\bibfnamefont {R.~B.-S.}\ \bibnamefont {Tsai}}, \bibinfo
  {author} {\bibfnamefont {P.}~\bibnamefont {Scholl}}, \bibinfo {author}
  {\bibfnamefont {T.~H.}\ \bibnamefont {Yoon}}, \bibinfo {author}
  {\bibfnamefont {J.}~\bibnamefont {Choi}}, \ and\ \bibinfo {author}
  {\bibfnamefont {M.}~\bibnamefont {Endres}},\ }\bibfield  {title} {\emph
  {\bibinfo {title} {Multi-ensemble metrology by programming local rotations
  with atom movements},\ }}\href {\doibase 10.1038/s41567-023-02323-w}
  {\bibfield  {journal} {\bibinfo  {journal} {Nat. Phys.}\ }\textbf {\bibinfo
  {volume} {20}},\ \bibinfo {pages} {195} (\bibinfo {year} {2024})}\BibitemShut
  {NoStop}%
\bibitem [{\citenamefont {Esteve}\ \emph {et~al.}(2006)\citenamefont {Esteve},
  \citenamefont {Trebbia}, \citenamefont {Schumm}, \citenamefont {Aspect},
  \citenamefont {Westbrook},\ and\ \citenamefont
  {Bouchoule}}]{esteve2006observations}%
  \BibitemOpen
  \bibfield  {author} {\bibinfo {author} {\bibfnamefont {J.}~\bibnamefont
  {Esteve}}, \bibinfo {author} {\bibfnamefont {J.-B.}\ \bibnamefont {Trebbia}},
  \bibinfo {author} {\bibfnamefont {T.}~\bibnamefont {Schumm}}, \bibinfo
  {author} {\bibfnamefont {A.}~\bibnamefont {Aspect}}, \bibinfo {author}
  {\bibfnamefont {C.~I.}\ \bibnamefont {Westbrook}}, \ and\ \bibinfo {author}
  {\bibfnamefont {I.}~\bibnamefont {Bouchoule}},\ }\bibfield  {title} {\emph
  {\bibinfo {title} {Observations of density fluctuations in an elongated bose
  gas: Ideal gas and quasicondensate regimes},\ }}\href {\doibase
  10.1103/PhysRevLett.96.130403} {\bibfield  {journal} {\bibinfo  {journal}
  {Phys. Rev. Lett.}\ }\textbf {\bibinfo {volume} {96}},\ \bibinfo {pages}
  {130403} (\bibinfo {year} {2006})}\BibitemShut {NoStop}%
\bibitem [{\citenamefont {Est{\`e}ve}\ \emph {et~al.}(2008)\citenamefont
  {Est{\`e}ve}, \citenamefont {Gross}, \citenamefont {Weller}, \citenamefont
  {Giovanazzi},\ and\ \citenamefont {Oberthaler}}]{esteve2008squeezing}%
  \BibitemOpen
  \bibfield  {author} {\bibinfo {author} {\bibfnamefont {J.}~\bibnamefont
  {Est{\`e}ve}}, \bibinfo {author} {\bibfnamefont {C.}~\bibnamefont {Gross}},
  \bibinfo {author} {\bibfnamefont {A.}~\bibnamefont {Weller}}, \bibinfo
  {author} {\bibfnamefont {S.}~\bibnamefont {Giovanazzi}}, \ and\ \bibinfo
  {author} {\bibfnamefont {M.~K.}\ \bibnamefont {Oberthaler}},\ }\bibfield
  {title} {\emph {\bibinfo {title} {Squeezing and entanglement in a
  bose--einstein condensate},\ }}\href {\doibase 10.1038/nature07332}
  {\bibfield  {journal} {\bibinfo  {journal} {Nature}\ }\textbf {\bibinfo
  {volume} {455}},\ \bibinfo {pages} {1216} (\bibinfo {year}
  {2008})}\BibitemShut {NoStop}%
\bibitem [{\citenamefont {Boixo}\ \emph {et~al.}(2007)\citenamefont {Boixo},
  \citenamefont {Flammia}, \citenamefont {Caves},\ and\ \citenamefont
  {Geremia}}]{boixo2007generalized}%
  \BibitemOpen
  \bibfield  {author} {\bibinfo {author} {\bibfnamefont {S.}~\bibnamefont
  {Boixo}}, \bibinfo {author} {\bibfnamefont {S.~T.}\ \bibnamefont {Flammia}},
  \bibinfo {author} {\bibfnamefont {C.~M.}\ \bibnamefont {Caves}}, \ and\
  \bibinfo {author} {\bibfnamefont {J.}~\bibnamefont {Geremia}},\ }\bibfield
  {title} {\emph {\bibinfo {title} {Generalized limits for single-parameter
  quantum estimation},\ }}\href {\doibase 10.1103/PhysRevLett.98.090401}
  {\bibfield  {journal} {\bibinfo  {journal} {Phys. Rev. Lett.}\ }\textbf
  {\bibinfo {volume} {98}},\ \bibinfo {pages} {090401} (\bibinfo {year}
  {2007})}\BibitemShut {NoStop}%
\bibitem [{\citenamefont {Ritboon}\ \emph {et~al.}(2022)\citenamefont
  {Ritboon}, \citenamefont {Slodi{\v{c}}ka},\ and\ \citenamefont
  {Filip}}]{ritboon2022sequential}%
  \BibitemOpen
  \bibfield  {author} {\bibinfo {author} {\bibfnamefont {A.}~\bibnamefont
  {Ritboon}}, \bibinfo {author} {\bibfnamefont {L.}~\bibnamefont
  {Slodi{\v{c}}ka}}, \ and\ \bibinfo {author} {\bibfnamefont {R.}~\bibnamefont
  {Filip}},\ }\bibfield  {title} {\emph {\bibinfo {title} {Sequential phonon
  measurements of atomic motion},\ }}\href {\doibase 10.1088/2058-9565/ac3c52}
  {\bibfield  {journal} {\bibinfo  {journal} {Quantum Sci. Technol.}\ }\textbf
  {\bibinfo {volume} {7}},\ \bibinfo {pages} {015023} (\bibinfo {year}
  {2022})}\BibitemShut {NoStop}%
\bibitem [{\citenamefont {Yang}\ \emph {et~al.}(2022)\citenamefont {Yang},
  \citenamefont {Pang}, \citenamefont {Chen}, \citenamefont {Jordan},\ and\
  \citenamefont {Del~Campo}}]{yang2022variational}%
  \BibitemOpen
  \bibfield  {author} {\bibinfo {author} {\bibfnamefont {J.}~\bibnamefont
  {Yang}}, \bibinfo {author} {\bibfnamefont {S.}~\bibnamefont {Pang}}, \bibinfo
  {author} {\bibfnamefont {Z.}~\bibnamefont {Chen}}, \bibinfo {author}
  {\bibfnamefont {A.~N.}\ \bibnamefont {Jordan}}, \ and\ \bibinfo {author}
  {\bibfnamefont {A.}~\bibnamefont {Del~Campo}},\ }\bibfield  {title} {\emph
  {\bibinfo {title} {Variational principle for optimal quantum controls in
  quantum metrology},\ }}\href {\doibase 10.1103/PhysRevLett.128.160505}
  {\bibfield  {journal} {\bibinfo  {journal} {Phys. Rev. Lett.}\ }\textbf
  {\bibinfo {volume} {128}},\ \bibinfo {pages} {160505} (\bibinfo {year}
  {2022})}\BibitemShut {NoStop}%
\bibitem [{\citenamefont {Xia}\ \emph {et~al.}(2024)\citenamefont {Xia},
  \citenamefont {Huang}, \citenamefont {Li}, \citenamefont {Luo},\ and\
  \citenamefont {Zeng}}]{xia2023nanoradian}%
  \BibitemOpen
  \bibfield  {author} {\bibinfo {author} {\bibfnamefont {B.}~\bibnamefont
  {Xia}}, \bibinfo {author} {\bibfnamefont {J.}~\bibnamefont {Huang}}, \bibinfo
  {author} {\bibfnamefont {H.}~\bibnamefont {Li}}, \bibinfo {author}
  {\bibfnamefont {Z.}~\bibnamefont {Luo}}, \ and\ \bibinfo {author}
  {\bibfnamefont {G.}~\bibnamefont {Zeng}},\ }\bibfield  {title} {\emph
  {\bibinfo {title} {Nanoradian-scale precision in light rotation measurement
  via indefinite quantum dynamics},\ }}\href {\doibase 10.1126/sciadv.adm8524}
  {\bibfield  {journal} {\bibinfo  {journal} {Sci. Adv.}\ }\textbf {\bibinfo
  {volume} {10}},\ \bibinfo {pages} {eadm8524} (\bibinfo {year}
  {2024})}\BibitemShut {NoStop}%
\bibitem [{\citenamefont {Fan}\ and\ \citenamefont
  {Pang}(2024)}]{fan2024achieving}%
  \BibitemOpen
  \bibfield  {author} {\bibinfo {author} {\bibfnamefont {J.}~\bibnamefont
  {Fan}}\ and\ \bibinfo {author} {\bibfnamefont {S.}~\bibnamefont {Pang}},\
  }\bibfield  {title} {\emph {\bibinfo {title} {Achieving {H}eisenberg scaling
  by probe-ancilla interaction in quantum metrology},\ }}\href {\doibase
  10.1103/PhysRevA.110.062406} {\bibfield  {journal} {\bibinfo  {journal}
  {Phys. Rev. A}\ }\textbf {\bibinfo {volume} {110}},\ \bibinfo {pages}
  {062406} (\bibinfo {year} {2024})}\BibitemShut {NoStop}%
\bibitem [{\citenamefont {Chen}\ and\ \citenamefont
  {Jing}(2024)}]{chen2024qubit}%
  \BibitemOpen
  \bibfield  {author} {\bibinfo {author} {\bibfnamefont {P.}~\bibnamefont
  {Chen}}\ and\ \bibinfo {author} {\bibfnamefont {J.}~\bibnamefont {Jing}},\
  }\bibfield  {title} {\emph {\bibinfo {title} {Qubit-assisted quantum
  metrology under a time-reversal strategy},\ }}\href {\doibase
  10.1103/PhysRevA.110.062425} {\bibfield  {journal} {\bibinfo  {journal}
  {Phys. Rev. A}\ }\textbf {\bibinfo {volume} {110}},\ \bibinfo {pages}
  {062425} (\bibinfo {year} {2024})}\BibitemShut {NoStop}%
\bibitem [{\citenamefont {Chen}\ and\ \citenamefont
  {Jing}(2025)}]{chen2025achieving}%
  \BibitemOpen
  \bibfield  {author} {\bibinfo {author} {\bibfnamefont {P.}~\bibnamefont
  {Chen}}\ and\ \bibinfo {author} {\bibfnamefont {J.}~\bibnamefont {Jing}},\
  }\bibfield  {title} {\emph {\bibinfo {title} {Achieving the heisenberg limit
  of metrology via measurement on an ancillary qubit},\ }}\href {\doibase
  10.1103/f79z-vjsb} {\bibfield  {journal} {\bibinfo  {journal} {Phys. Rev. A}\
  }\textbf {\bibinfo {volume} {112}},\ \bibinfo {pages} {032416} (\bibinfo
  {year} {2025})}\BibitemShut {NoStop}%
\bibitem [{\citenamefont {McCormick}\ \emph {et~al.}(2019)\citenamefont
  {McCormick}, \citenamefont {Keller}, \citenamefont {Burd}, \citenamefont
  {Wineland}, \citenamefont {Wilson},\ and\ \citenamefont
  {Leibfried}}]{mccormick2019quantum}%
  \BibitemOpen
  \bibfield  {author} {\bibinfo {author} {\bibfnamefont {K.~C.}\ \bibnamefont
  {McCormick}}, \bibinfo {author} {\bibfnamefont {J.}~\bibnamefont {Keller}},
  \bibinfo {author} {\bibfnamefont {S.~C.}\ \bibnamefont {Burd}}, \bibinfo
  {author} {\bibfnamefont {D.~J.}\ \bibnamefont {Wineland}}, \bibinfo {author}
  {\bibfnamefont {A.~C.}\ \bibnamefont {Wilson}}, \ and\ \bibinfo {author}
  {\bibfnamefont {D.}~\bibnamefont {Leibfried}},\ }\bibfield  {title} {\emph
  {\bibinfo {title} {Quantum-enhanced sensing of a single-ion mechanical
  oscillator},\ }}\href {\doibase 10.1038/s41586-019-1421-y} {\bibfield
  {journal} {\bibinfo  {journal} {Nature}\ }\textbf {\bibinfo {volume} {572}},\
  \bibinfo {pages} {86} (\bibinfo {year} {2019})}\BibitemShut {NoStop}%
\bibitem [{\citenamefont {Deng}\ \emph {et~al.}(2024)\citenamefont {Deng},
  \citenamefont {Li}, \citenamefont {Chen}, \citenamefont {Ni}, \citenamefont
  {Cai}, \citenamefont {Mai}, \citenamefont {Zhang}, \citenamefont {Zheng},
  \citenamefont {Yu}, \citenamefont {Zou} \emph {et~al.}}]{deng2024quantum}%
  \BibitemOpen
  \bibfield  {author} {\bibinfo {author} {\bibfnamefont {X.}~\bibnamefont
  {Deng}}, \bibinfo {author} {\bibfnamefont {S.}~\bibnamefont {Li}}, \bibinfo
  {author} {\bibfnamefont {Z.-J.}\ \bibnamefont {Chen}}, \bibinfo {author}
  {\bibfnamefont {Z.}~\bibnamefont {Ni}}, \bibinfo {author} {\bibfnamefont
  {Y.}~\bibnamefont {Cai}}, \bibinfo {author} {\bibfnamefont {J.}~\bibnamefont
  {Mai}}, \bibinfo {author} {\bibfnamefont {L.}~\bibnamefont {Zhang}}, \bibinfo
  {author} {\bibfnamefont {P.}~\bibnamefont {Zheng}}, \bibinfo {author}
  {\bibfnamefont {H.}~\bibnamefont {Yu}}, \bibinfo {author} {\bibfnamefont
  {C.-L.}\ \bibnamefont {Zou}},  \emph {et~al.},\ }\bibfield  {title} {\emph
  {\bibinfo {title} {Quantum-enhanced metrology with large fock states},\
  }}\href {\doibase 10.1038/s41567-024-02619-5} {\bibfield  {journal} {\bibinfo
   {journal} {Nat. Phys.}\ }\textbf {\bibinfo {volume} {20}},\ \bibinfo {pages}
  {1874} (\bibinfo {year} {2024})}\BibitemShut {NoStop}%
\bibitem [{\citenamefont {Pan}\ \emph {et~al.}(2025)\citenamefont {Pan},
  \citenamefont {Krisnanda}, \citenamefont {Duina}, \citenamefont {Park},
  \citenamefont {Song}, \citenamefont {Fontaine}, \citenamefont {Copetudo},
  \citenamefont {Filip},\ and\ \citenamefont {Gao}}]{pan2025realization}%
  \BibitemOpen
  \bibfield  {author} {\bibinfo {author} {\bibfnamefont {X.}~\bibnamefont
  {Pan}}, \bibinfo {author} {\bibfnamefont {T.}~\bibnamefont {Krisnanda}},
  \bibinfo {author} {\bibfnamefont {A.}~\bibnamefont {Duina}}, \bibinfo
  {author} {\bibfnamefont {K.}~\bibnamefont {Park}}, \bibinfo {author}
  {\bibfnamefont {P.}~\bibnamefont {Song}}, \bibinfo {author} {\bibfnamefont
  {C.~Y.}\ \bibnamefont {Fontaine}}, \bibinfo {author} {\bibfnamefont
  {A.}~\bibnamefont {Copetudo}}, \bibinfo {author} {\bibfnamefont
  {R.}~\bibnamefont {Filip}}, \ and\ \bibinfo {author} {\bibfnamefont {Y.~Y.}\
  \bibnamefont {Gao}},\ }\bibfield  {title} {\emph {\bibinfo {title}
  {Realization of versatile and effective quantum metrology using a single
  bosonic mode},\ }}\href {\doibase 10.1103/PRXQuantum.6.010304} {\bibfield
  {journal} {\bibinfo  {journal} {PRX Quantum}\ }\textbf {\bibinfo {volume}
  {6}},\ \bibinfo {pages} {010304} (\bibinfo {year} {2025})}\BibitemShut
  {NoStop}%
\bibitem [{\citenamefont {Helstrom}(1969)}]{helstrom1969quantum}%
  \BibitemOpen
  \bibfield  {author} {\bibinfo {author} {\bibfnamefont {C.~W.}\ \bibnamefont
  {Helstrom}},\ }\bibfield  {title} {\emph {\bibinfo {title} {Quantum detection
  and estimation theory},\ }}\href {\doibase 10.1007/BF01007479} {\bibfield
  {journal} {\bibinfo  {journal} {J. Stat. Phys.}\ }\textbf {\bibinfo {volume}
  {1}},\ \bibinfo {pages} {231} (\bibinfo {year} {1969})}\BibitemShut {NoStop}%
\bibitem [{\citenamefont {Holevo}(2011)}]{holevo2011probabilistic}%
  \BibitemOpen
  \bibfield  {author} {\bibinfo {author} {\bibfnamefont {A.~S.}\ \bibnamefont
  {Holevo}},\ }\href@noop {} {\emph {\bibinfo {title} {{Probabilistic and
  Statistical Aspects of Quantum Theory}}}},\ Vol.~\bibinfo {volume} {1}\
  (\bibinfo  {publisher} {Springer},\ \bibinfo {address} {Pisa},\ \bibinfo
  {year} {2011})\BibitemShut {NoStop}%
\bibitem [{\citenamefont {Braunstein}\ and\ \citenamefont
  {Caves}(1994)}]{braunstein1994statistical}%
  \BibitemOpen
  \bibfield  {author} {\bibinfo {author} {\bibfnamefont {S.~L.}\ \bibnamefont
  {Braunstein}}\ and\ \bibinfo {author} {\bibfnamefont {C.~M.}\ \bibnamefont
  {Caves}},\ }\bibfield  {title} {\emph {\bibinfo {title} {Statistical distance
  and the geometry of quantum states},\ }}\href {\doibase
  10.1103/PhysRevLett.72.3439} {\bibfield  {journal} {\bibinfo  {journal}
  {Phys. Rev. Lett.}\ }\textbf {\bibinfo {volume} {72}},\ \bibinfo {pages}
  {3439} (\bibinfo {year} {1994})}\BibitemShut {NoStop}%
\bibitem [{\citenamefont {Braunstein}\ \emph {et~al.}(1996)\citenamefont
  {Braunstein}, \citenamefont {Caves},\ and\ \citenamefont
  {Milburn}}]{braunstein1996generalized}%
  \BibitemOpen
  \bibfield  {author} {\bibinfo {author} {\bibfnamefont {S.~L.}\ \bibnamefont
  {Braunstein}}, \bibinfo {author} {\bibfnamefont {C.~M.}\ \bibnamefont
  {Caves}}, \ and\ \bibinfo {author} {\bibfnamefont {G.~J.}\ \bibnamefont
  {Milburn}},\ }\bibfield  {title} {\emph {\bibinfo {title} {Generalized
  uncertainty relations: theory, examples, and lorentz invariance},\ }}\href
  {\doibase https://doi.org/10.1006/aphy.1996.0040} {\bibfield  {journal}
  {\bibinfo  {journal} {Ann. Phys.}\ }\textbf {\bibinfo {volume} {247}},\
  \bibinfo {pages} {135} (\bibinfo {year} {1996})}\BibitemShut {NoStop}%
\bibitem [{\citenamefont {Zhang}\ \emph {et~al.}(2013)\citenamefont {Zhang},
  \citenamefont {Li}, \citenamefont {Yang},\ and\ \citenamefont
  {Jin}}]{zhang2013quantum}%
  \BibitemOpen
  \bibfield  {author} {\bibinfo {author} {\bibfnamefont {Y.-M.}\ \bibnamefont
  {Zhang}}, \bibinfo {author} {\bibfnamefont {X.-W.}\ \bibnamefont {Li}},
  \bibinfo {author} {\bibfnamefont {W.}~\bibnamefont {Yang}}, \ and\ \bibinfo
  {author} {\bibfnamefont {G.-R.}\ \bibnamefont {Jin}},\ }\bibfield  {title}
  {\emph {\bibinfo {title} {Quantum {Fisher} information of entangled coherent
  states in the presence of photon loss},\ }}\href {\doibase
  10.1103/PhysRevA.88.043832} {\bibfield  {journal} {\bibinfo  {journal} {Phys.
  Rev. A}\ }\textbf {\bibinfo {volume} {88}},\ \bibinfo {pages} {043832}
  (\bibinfo {year} {2013})}\BibitemShut {NoStop}%
\bibitem [{\citenamefont {Liu}\ \emph {et~al.}(2014)\citenamefont {Liu},
  \citenamefont {Jing}, \citenamefont {Zhong},\ and\ \citenamefont
  {Wang}}]{liu2014quantum}%
  \BibitemOpen
  \bibfield  {author} {\bibinfo {author} {\bibfnamefont {J.}~\bibnamefont
  {Liu}}, \bibinfo {author} {\bibfnamefont {X.-X.}\ \bibnamefont {Jing}},
  \bibinfo {author} {\bibfnamefont {W.}~\bibnamefont {Zhong}}, \ and\ \bibinfo
  {author} {\bibfnamefont {X.-G.}\ \bibnamefont {Wang}},\ }\bibfield  {title}
  {\emph {\bibinfo {title} {Quantum {Fisher} information for density matrices
  with arbitrary ranks},\ }}\href {\doibase 10.1088/0253-6102/61/1/08}
  {\bibfield  {journal} {\bibinfo  {journal} {Commun. Theor. Phys.}\ }\textbf
  {\bibinfo {volume} {61}},\ \bibinfo {pages} {45} (\bibinfo {year}
  {2014})}\BibitemShut {NoStop}%
\bibitem [{\citenamefont {Cahen}\ and\ \citenamefont
  {Chabert}(1997)}]{cahen1997integer}%
  \BibitemOpen
  \bibfield  {author} {\bibinfo {author} {\bibfnamefont {P.-J.}\ \bibnamefont
  {Cahen}}\ and\ \bibinfo {author} {\bibfnamefont {J.-L.}\ \bibnamefont
  {Chabert}},\ }\href@noop {} {\emph {\bibinfo {title} {{Integer-valued
  Polynomials}}}},\ \bibinfo {series} {Mathematical Surveys and Monographs},
  Vol.~\bibinfo {volume} {48}\ (\bibinfo  {publisher} {American Mathematical
  Society},\ \bibinfo {address} {Providence, RI},\ \bibinfo {year}
  {1997})\BibitemShut {NoStop}%
\bibitem [{\citenamefont {Wang}\ \emph {et~al.}(2015)\citenamefont {Wang},
  \citenamefont {Yuan}, \citenamefont {Huang}, \citenamefont {Rong},
  \citenamefont {Wang}, \citenamefont {Xu}, \citenamefont {Duan}, \citenamefont
  {Ju}, \citenamefont {Shi},\ and\ \citenamefont {Du}}]{wang2015high}%
  \BibitemOpen
  \bibfield  {author} {\bibinfo {author} {\bibfnamefont {P.}~\bibnamefont
  {Wang}}, \bibinfo {author} {\bibfnamefont {Z.}~\bibnamefont {Yuan}}, \bibinfo
  {author} {\bibfnamefont {P.}~\bibnamefont {Huang}}, \bibinfo {author}
  {\bibfnamefont {X.}~\bibnamefont {Rong}}, \bibinfo {author} {\bibfnamefont
  {M.}~\bibnamefont {Wang}}, \bibinfo {author} {\bibfnamefont {X.}~\bibnamefont
  {Xu}}, \bibinfo {author} {\bibfnamefont {C.}~\bibnamefont {Duan}}, \bibinfo
  {author} {\bibfnamefont {C.}~\bibnamefont {Ju}}, \bibinfo {author}
  {\bibfnamefont {F.}~\bibnamefont {Shi}}, \ and\ \bibinfo {author}
  {\bibfnamefont {J.}~\bibnamefont {Du}},\ }\bibfield  {title} {\emph {\bibinfo
  {title} {High-resolution vector microwave magnetometry based on solid-state
  spins in diamond},\ }}\href {\doibase 10.1038/ncomms7631} {\bibfield
  {journal} {\bibinfo  {journal} {Nat. commun.}\ }\textbf {\bibinfo {volume}
  {6}},\ \bibinfo {pages} {6631} (\bibinfo {year} {2015})}\BibitemShut
  {NoStop}%
\bibitem [{\citenamefont {Boto}\ \emph {et~al.}(2018)\citenamefont {Boto},
  \citenamefont {Holmes}, \citenamefont {Leggett}, \citenamefont {Roberts},
  \citenamefont {Shah}, \citenamefont {Meyer}, \citenamefont {Mu{\~n}oz},
  \citenamefont {Mullinger}, \citenamefont {Tierney}, \citenamefont {Bestmann}
  \emph {et~al.}}]{boto2018moving}%
  \BibitemOpen
  \bibfield  {author} {\bibinfo {author} {\bibfnamefont {E.}~\bibnamefont
  {Boto}}, \bibinfo {author} {\bibfnamefont {N.}~\bibnamefont {Holmes}},
  \bibinfo {author} {\bibfnamefont {J.}~\bibnamefont {Leggett}}, \bibinfo
  {author} {\bibfnamefont {G.}~\bibnamefont {Roberts}}, \bibinfo {author}
  {\bibfnamefont {V.}~\bibnamefont {Shah}}, \bibinfo {author} {\bibfnamefont
  {S.~S.}\ \bibnamefont {Meyer}}, \bibinfo {author} {\bibfnamefont {L.~D.}\
  \bibnamefont {Mu{\~n}oz}}, \bibinfo {author} {\bibfnamefont {K.~J.}\
  \bibnamefont {Mullinger}}, \bibinfo {author} {\bibfnamefont {T.~M.}\
  \bibnamefont {Tierney}}, \bibinfo {author} {\bibfnamefont {S.}~\bibnamefont
  {Bestmann}},  \emph {et~al.},\ }\bibfield  {title} {\emph {\bibinfo {title}
  {Moving magnetoencephalography towards real-world applications with a
  wearable system},\ }}\href {\doibase doi.org/10.1038/nature26147} {\bibfield
  {journal} {\bibinfo  {journal} {Nature}\ }\textbf {\bibinfo {volume} {555}},\
  \bibinfo {pages} {657} (\bibinfo {year} {2018})}\BibitemShut {NoStop}%
\bibitem [{\citenamefont {Xie}\ \emph {et~al.}(2021)\citenamefont {Xie},
  \citenamefont {Zhao}, \citenamefont {Kong}, \citenamefont {Ma}, \citenamefont
  {Wang}, \citenamefont {Ye}, \citenamefont {Yu}, \citenamefont {Yang},
  \citenamefont {Xu}, \citenamefont {Wang} \emph {et~al.}}]{xie2021beating}%
  \BibitemOpen
  \bibfield  {author} {\bibinfo {author} {\bibfnamefont {T.}~\bibnamefont
  {Xie}}, \bibinfo {author} {\bibfnamefont {Z.}~\bibnamefont {Zhao}}, \bibinfo
  {author} {\bibfnamefont {X.}~\bibnamefont {Kong}}, \bibinfo {author}
  {\bibfnamefont {W.}~\bibnamefont {Ma}}, \bibinfo {author} {\bibfnamefont
  {M.}~\bibnamefont {Wang}}, \bibinfo {author} {\bibfnamefont {X.}~\bibnamefont
  {Ye}}, \bibinfo {author} {\bibfnamefont {P.}~\bibnamefont {Yu}}, \bibinfo
  {author} {\bibfnamefont {Z.}~\bibnamefont {Yang}}, \bibinfo {author}
  {\bibfnamefont {S.}~\bibnamefont {Xu}}, \bibinfo {author} {\bibfnamefont
  {P.}~\bibnamefont {Wang}},  \emph {et~al.},\ }\bibfield  {title} {\emph
  {\bibinfo {title} {Beating the standard quantum limit under ambient
  conditions with solid-state spins},\ }}\href {\doibase
  10.1126/sciadv.abg9204} {\bibfield  {journal} {\bibinfo  {journal} {Sci.
  Adv.}\ }\textbf {\bibinfo {volume} {7}},\ \bibinfo {pages} {eabg9204}
  (\bibinfo {year} {2021})}\BibitemShut {NoStop}%
\bibitem [{\citenamefont {Montenegro}\ \emph {et~al.}(2022)\citenamefont
  {Montenegro}, \citenamefont {Jones}, \citenamefont {Bose},\ and\
  \citenamefont {Bayat}}]{montenegro2022sequential}%
  \BibitemOpen
  \bibfield  {author} {\bibinfo {author} {\bibfnamefont {V.}~\bibnamefont
  {Montenegro}}, \bibinfo {author} {\bibfnamefont {G.~S.}\ \bibnamefont
  {Jones}}, \bibinfo {author} {\bibfnamefont {S.}~\bibnamefont {Bose}}, \ and\
  \bibinfo {author} {\bibfnamefont {A.}~\bibnamefont {Bayat}},\ }\bibfield
  {title} {\emph {\bibinfo {title} {Sequential measurements for
  quantum-enhanced magnetometry in spin chain probes},\ }}\href {\doibase
  10.1103/PhysRevLett.129.120503} {\bibfield  {journal} {\bibinfo  {journal}
  {Phys. Rev. Lett.}\ }\textbf {\bibinfo {volume} {129}},\ \bibinfo {pages}
  {120503} (\bibinfo {year} {2022})}\BibitemShut {NoStop}%
\bibitem [{\citenamefont {Yu}\ \emph {et~al.}(2025)\citenamefont {Yu},
  \citenamefont {Zhu}, \citenamefont {Zhang}, \citenamefont {Jing},
  \citenamefont {Wang}, \citenamefont {Chen}, \citenamefont {Xie},
  \citenamefont {Rong},\ and\ \citenamefont {Du}}]{yu2025experimental}%
  \BibitemOpen
  \bibfield  {author} {\bibinfo {author} {\bibfnamefont {Z.}~\bibnamefont
  {Yu}}, \bibinfo {author} {\bibfnamefont {Y.}~\bibnamefont {Zhu}}, \bibinfo
  {author} {\bibfnamefont {W.}~\bibnamefont {Zhang}}, \bibinfo {author}
  {\bibfnamefont {K.}~\bibnamefont {Jing}}, \bibinfo {author} {\bibfnamefont
  {S.}~\bibnamefont {Wang}}, \bibinfo {author} {\bibfnamefont {C.}~\bibnamefont
  {Chen}}, \bibinfo {author} {\bibfnamefont {Y.}~\bibnamefont {Xie}}, \bibinfo
  {author} {\bibfnamefont {X.}~\bibnamefont {Rong}}, \ and\ \bibinfo {author}
  {\bibfnamefont {J.}~\bibnamefont {Du}},\ }\bibfield  {title} {\emph {\bibinfo
  {title} {Experimental demonstration of a diamond quantum vector magnetometer
  for deep-sea applications},\ }}\href {\doibase 10.1093/nsr/nwae478}
  {\bibfield  {journal} {\bibinfo  {journal} {Natl. Sci. Rev.}\ }\textbf
  {\bibinfo {volume} {12}},\ \bibinfo {pages} {nwae478} (\bibinfo {year}
  {2025})}\BibitemShut {NoStop}%
\bibitem [{\citenamefont {Le~Sage}\ \emph {et~al.}(2012)\citenamefont
  {Le~Sage}, \citenamefont {Pham}, \citenamefont {Bar-Gill}, \citenamefont
  {Belthangady}, \citenamefont {Lukin}, \citenamefont {Yacoby},\ and\
  \citenamefont {Walsworth}}]{le2012efficient}%
  \BibitemOpen
  \bibfield  {author} {\bibinfo {author} {\bibfnamefont {D.}~\bibnamefont
  {Le~Sage}}, \bibinfo {author} {\bibfnamefont {L.~M.}\ \bibnamefont {Pham}},
  \bibinfo {author} {\bibfnamefont {N.}~\bibnamefont {Bar-Gill}}, \bibinfo
  {author} {\bibfnamefont {C.}~\bibnamefont {Belthangady}}, \bibinfo {author}
  {\bibfnamefont {M.~D.}\ \bibnamefont {Lukin}}, \bibinfo {author}
  {\bibfnamefont {A.}~\bibnamefont {Yacoby}}, \ and\ \bibinfo {author}
  {\bibfnamefont {R.~L.}\ \bibnamefont {Walsworth}},\ }\bibfield  {title}
  {\emph {\bibinfo {title} {Efficient photon detection from color centers in a
  diamond optical waveguide},\ }}\href {\doibase 10.1103/PhysRevB.85.121202}
  {\bibfield  {journal} {\bibinfo  {journal} {Phys. Rev. B}\ }\textbf {\bibinfo
  {volume} {85}},\ \bibinfo {pages} {121202} (\bibinfo {year}
  {2012})}\BibitemShut {NoStop}%
\bibitem [{\citenamefont {Yurke}\ \emph {et~al.}(1986)\citenamefont {Yurke},
  \citenamefont {McCall},\ and\ \citenamefont {Klauder}}]{yurke19862}%
  \BibitemOpen
  \bibfield  {author} {\bibinfo {author} {\bibfnamefont {B.}~\bibnamefont
  {Yurke}}, \bibinfo {author} {\bibfnamefont {S.~L.}\ \bibnamefont {McCall}}, \
  and\ \bibinfo {author} {\bibfnamefont {J.~R.}\ \bibnamefont {Klauder}},\
  }\bibfield  {title} {\emph {\bibinfo {title} {{SU}(2) and {SU}(1,1)
  interferometers},\ }}\href {\doibase 10.1103/PhysRevA.33.4033} {\bibfield
  {journal} {\bibinfo  {journal} {Phys. Rev. A}\ }\textbf {\bibinfo {volume}
  {33}},\ \bibinfo {pages} {4033} (\bibinfo {year} {1986})}\BibitemShut
  {NoStop}%
\bibitem [{\citenamefont {Pezz\`e}\ \emph {et~al.}(2017)\citenamefont
  {Pezz\`e}, \citenamefont {Ciampini}, \citenamefont {Spagnolo}, \citenamefont
  {Humphreys}, \citenamefont {Datta}, \citenamefont {Walmsley}, \citenamefont
  {Barbieri}, \citenamefont {Sciarrino},\ and\ \citenamefont
  {Smerzi}}]{pezze2017optimal}%
  \BibitemOpen
  \bibfield  {author} {\bibinfo {author} {\bibfnamefont {L.}~\bibnamefont
  {Pezz\`e}}, \bibinfo {author} {\bibfnamefont {M.~A.}\ \bibnamefont
  {Ciampini}}, \bibinfo {author} {\bibfnamefont {N.}~\bibnamefont {Spagnolo}},
  \bibinfo {author} {\bibfnamefont {P.~C.}\ \bibnamefont {Humphreys}}, \bibinfo
  {author} {\bibfnamefont {A.}~\bibnamefont {Datta}}, \bibinfo {author}
  {\bibfnamefont {I.~A.}\ \bibnamefont {Walmsley}}, \bibinfo {author}
  {\bibfnamefont {M.}~\bibnamefont {Barbieri}}, \bibinfo {author}
  {\bibfnamefont {F.}~\bibnamefont {Sciarrino}}, \ and\ \bibinfo {author}
  {\bibfnamefont {A.}~\bibnamefont {Smerzi}},\ }\bibfield  {title} {\emph
  {\bibinfo {title} {Optimal measurements for simultaneous quantum estimation
  of multiple phases},\ }}\href {\doibase 10.1103/PhysRevLett.119.130504}
  {\bibfield  {journal} {\bibinfo  {journal} {Phys. Rev. Lett.}\ }\textbf
  {\bibinfo {volume} {119}},\ \bibinfo {pages} {130504} (\bibinfo {year}
  {2017})}\BibitemShut {NoStop}%
\end{thebibliography}%

\end{document}